\documentclass[12pt]{article}
\pdfoutput=1
\usepackage{latexsym, graphicx,cite}
\usepackage{amsmath}
\usepackage{amssymb}
\usepackage{amsthm}
\usepackage{tikz-feynman}
\usepackage{caption}
\usepackage{subcaption}
\usepackage{dsfont}

\allowdisplaybreaks

\usepackage[top = 1. in,bottom = 0.93 in, left = 1 in, right=1 in]{geometry}

 \newcommand{\arXiv}[1]{\href{http://www.arXiv.org/abs/#1}{arXiv:#1}}
\usepackage[colorlinks=true, linkcolor=blue, bookmarks=true]{hyperref}

\makeatletter
\renewcommand\section{\@startsection {section}{1}{\z@}%
                  {-3.5ex \@plus -1ex \@minus -.2ex}
                  {2.3ex \@plus.2ex}%
                  {\normalfont\large\bfseries}}
\renewcommand\subsection{\@startsection{subsection}{2}{\z@}%
                   {-3.25ex\@plus -1ex \@minus -.2ex}%
                   {1.5ex \@plus .2ex}%
                   {\normalfont\bfseries}}
\makeatother

\newcommand{\beq}{\begin{equation}}
\newcommand{\eeq}{\end{equation}}
\newcommand{\ber}{\begin{array}}
\newcommand{\eer}{\end{array}}
\newcommand{\del}{\partial}

\newcommand{\dsty}{\displaystyle}

\newcommand{\de}{\delta}
\newcommand{\ena}{\end{eqnarray}}
\newcommand{\beqa}{\begin{eqnarray}}
\newcommand{\eeqa}{\end{eqnarray}}
\newcommand{\bea}{\begin{eqnarray}}
\newcommand{\eea}{\end{eqnarray}}

\newcommand{\ER}{Erd\H{o}s-R\'enyi }
\newcommand{\kk}{c}

\newcommand{\pdet}{\text{det}' }
\newcommand{\sumph}{{\textstyle\sum_i}\phi_i}

\renewcommand{\Re}{\operatorname{Re}}
\renewcommand{\Im}{\operatorname{Im}}

\begin{document}
\begin{titlepage}
\begin{flushright}
\phantom{arXiv:yymm.nnnn}
\end{flushright}
\vspace{-5mm}
\begin{center}
{\huge\bf Resistance distance distribution in\vspace{3mm}\\ large sparse random graphs}\\
\vskip 15mm
{\large Pawat Akara-pipattana$^{a,b}$, Thiparat Chotibut$^{a,b}$ and Oleg Evnin$^{a,c}$}
\vskip 7mm
{\em $^a$ Department of Physics, Faculty of Science, Chulalongkorn University,
Bangkok, Thailand}
\vskip 3mm
{\em $^b$ Chula Intelligent and Complex Systems, Faculty of Science,\\
Chulalongkorn University, Bangkok, Thailand}
\vskip 3mm
{\em $^c$ Theoretische Natuurkunde, Vrije Universiteit Brussel and\\
The International Solvay Institutes, Brussels, Belgium}
\vskip 7mm
{\small\noindent {\tt akarapawat@gmail.com, thiparatc@gmail.com, oleg.evnin@gmail.com}}
\vskip 20mm
\end{center}
\begin{center}
{\bf ABSTRACT}\vspace{3mm}
\end{center}
We consider an Erd\H{o}s-R\'enyi random graph consisting of $N$ vertices connected by randomly and independently drawing an edge between every pair of them with probability $\kk/N$ so that at $N\to\infty$ one obtains a graph of finite mean degree $\kk$. In this regime, we study the distribution of resistance distances between the vertices of this graph and develop an auxiliary field representation for this quantity in the spirit of statistical field theory. Using this representation, a saddle point evaluation of the resistance distance distribution is possible at $N\to\infty$ in terms of an $1/\kk$ expansion. The leading order of this expansion captures the results of numerical simulations very well down to rather small values of $\kk$; for example, it recovers the empirical distribution at $\kk=4$ or 6 with an overlap of around 90\%. At large values of $\kk$, the distribution tends to a Gaussian of mean $2/\kk$ and standard deviation $\sqrt{2/\kk^3}$. At small values of $\kk$, the distribution is skewed toward larger values, as captured by our saddle point analysis,  and many fine features appear in addition to the main peak, including subleading peaks that can be traced back to resistance distances between vertices of specific low degrees and the rest of the graph. We develop a more refined saddle point scheme that extracts the corresponding degree-differentiated resistance distance distributions. We then use this approach to recover analytically the most apparent of the subleading peaks that originates from vertices of degree 1. Rather intuitively, this subleading peak turns out to be a copy of the main peak, shifted by one unit of resistance distance and scaled down by the probability for a vertex to have degree 1. We comment on a possible lack of smoothness in the true $N\to\infty$ distribution suggested by the numerics. 
\vfill

\end{titlepage}

\section{Introduction}

Of all the notions of distance that may be defined on graphs \cite{harary,dist}, perhaps the most evident one is given by the shortest path, or geodesic \cite{geo}, distance: the minimal number of edges one must traverse in order to transit from vertex $i$ to vertex $j$. While intuitive and visual, this notion of distance is limited in that it does not fully capture the ease or difficulty of reaching point $j$ from point $i$ by navigating the graph edges. It does not say whether there is only one path of minimal length or many such paths, whether these paths can be straightforwardly located, or whether alternative paths are considerably or only slightly longer.

A complementary viewpoint is provided by the {\it resistance distance} \cite{resdist1,resdist2,resdist3,resdist4,resdist5} defined by assigning resistances of 1 ohm to the graph edges and measuring the ordinary electric resistance between vertices $i$ and $j$. This quantity is naturally expressed through the inverse of the {\it graph Laplacian} and is therefore closely related \cite{commute} to diffusion and random walks on graphs \cite{lovasz}, measuring the time a random walker starting at point $i$ typically needs to reach point $j$, which leads to the alternative name {\it commute distance}. Resistance distances thus in principle capture properties of all possible paths, though there are important qualifications to this statement \cite{lostspace1}, see below. In view of these appealing properties, resistance distances have surfaced in research on subjects as diverse as theoretical physics \cite{phys1,phys2}, chemistry and bioinformatics \cite{chem1,chem1a,chem2,chem2a,chem3,chem4}, mathematical  graph theory \cite{graphinv,cacti1,cacti2}, data analysis and computer science \cite{cs1,cs2,cs3,cs4,cs5,cs6,cs7,cs8,cs9,cs10,cs11,cs12,cs13}. Studying resistance distance can also be useful for understanding nutrient transport in leaf vascular networks, as hydraulic conductance of laminar flows in this setting can be equivalently studied in terms of electrical conductance in resistor networks \cite{vasc1,vasc2,vasc3,vasc4}.

One important characterization of graph distances is given by how they behave in random graphs. For the shortest path distance, this question has been studied rather extensively \cite{diam1,diam2,short1,short2,short3,short4}.
By contrast, studies of the distribution of resistance distances have mostly been limited to concentration phenomena, as in \cite{lostspace1,resdistconc1,resdistconc2}. Indeed, under the assumption that the mean degree grows with the number of vertices (this growth may be very modest, for example, logarithmic), in the large graph limit, the resistance distance between vertices $i$ and $j$ of degrees $d_i$ and $d_j$ almost surely equals $1/d_i+1/d_j$,
the relevance of this number pointed out already in \cite{lovasz}. These concentration phenomena inspired the critique of resistance distance developed in \cite{lostspace1}, as the simple number $1/d_i+1/d_j$ cannot possibly hold any refined information on the path properties of graphs, and some modifications of the usual resistance distance definition have been proposed \cite{lostspace2} to address these shortcomings.\footnote{We
mention additionally that studies have been undertaken of systems where the resistance values assigned to the edges, rather than the underlying graph structure, are random \cite{randres}. The problem of choosing the edge resistance values on a given graph to minimize the overall average resistance has been considered in \cite{minimize}.} At the same time, concentration phenomena do not occur if the mean degree stays finite at $N\to\infty$ and the resistance distributions are not only nontrivial, but also ornately shaped. This is straightforwardly verified by simulating a finite mean degree \ER graph obtained by randomly and independently connecting $N$ points with probability $c/N$ per edge, where $N$ is large and the mean vertex degree $c$ is fixed.

Our goal in this paper is precisely to develop an analytic picture of resistance distance distributions in finite-mean-degree \ER graphs. (In other words, our focus is on large, sparse graphs with $c\ll N$.) While the motivations for studying resistance distances are broad and interdisciplinary, the methodology we employ is decidedly that of theoretical physics, and more specifically statistical field theory \cite{statFT, statFTneu}. We shall first develop an auxiliary field representation for the resistance distance distribution in the spirit of the field-theoretic Hubbard-Stratonovich transformation, which will convert the resistance distance  computation to the analysis of a statistical field theory on a complete graph with $N$ vertices. Then, in a manner common for large $N$ limits \cite{largeN} in statistical and quantum field theory, and as a particular realization of the `mean field' principles, we shall identify a saddle point that dominates the auxiliary field theory computation, recovering an estimate for the resistance distance distribution. Similar techniques have been applied to a variety of random graph problems in \cite{PN,spectrum,AC,corr}. (Earlier work applying more conventional probabilistic and combinatiorial methods to related spectral problems for random graphs can be found in \cite{BG}.) The level of  rigor in our treatment will likewise be typical of statistical field theory, and our approach is of empirical and heuristic nature: we shall identify saddle points that plausibly dominate the quantity of interest and develop estimates based on these saddle points, without attempting to rigorously control the accuracy of these estimates by analytic methods. The true judge of our endeavors is their agreement with numerical simulations that we shall systematically report.

Our study is organized as follows: first, in section~\ref{setup}, we shall provide the basic setup for the resistance distance distribution of \ER graphs and develop our auxiliary field representation. Then, in section~\ref{saddle}, we shall describe a saddle point treatment of this auxiliary field representation at leading order in $1/c$ in a simplified setup that ignores the fluctuations of the graph Laplacian determinant, and in section~\ref{sub}, give a full justification for the simplified analysis of section~\ref{saddle}. In section~\ref{deg1}, we shall focus on minor features of the resistance distance distribution not captured by our saddle point analysis and develop an analytic description of the simplest of these features. We shall conclude with a discussion and some tentative remarks on the roughness of the true resistance distance distribution at $N\to\infty$.

\section{Resistance distance distribution and its auxiliary field\\ representation}\label{setup}

As customary in random graph theory, we shall characterize graphs with $N$ vertices by their $N\times N$ {\it adjacency matrices} $A_{ij}$, which are taken to be symmetric with a vanishing diagonal. One assigns $1$ to $A_{ij}$ if there is an edge connecting vertices $i$ and $j$, and 0 otherwise.

To describe a large sparse \ER random graph in this language, we treat $A_{ij}$ with $i<j$ as independent random variables taking value 1 with probability $\kk/N$ and 0 with probability $1-\kk/N$. The expectation values of the vertex degrees
\beq
d_i\equiv \sum_{j=1}^N A_{ij}
\label{degree}
\eeq
are then $\kk(N-1)/N\approx \kk$ in the large $N$ limit. It is convenient for our purposes to recast this ensemble in the 
exponential random graph language \cite{newman} better adapted for statistical physics considerations. To each configuration $A_{ij}$ one assigns a Boltzmann weight $\exp[\ln\left(\frac{\kk}N\right)\sum_{i<j}A_{ij}]=(\kk/N)^{\sum_{i<j}A_{ij}}$, so that $\ln(N/\kk)$ plays the role of inverse temperature conjugate to the total number of graph edges. Then, the expectation value of any observable $F({\bf A})$ is given by
\beq
\langle F \rangle =\frac1Z\sum_{\{{\bf A}\}}F({\bf A})\left(\frac\kk N\right)^{\sum_{i<j}A_{ij}},
\eeq
with the partition function $Z\equiv \sum_{\{{\bf A}\}}\left(\frac\kk N\right)^{\sum_{i<j}A_{ij}}\approx e^{\kk(N-1)/{2}}$, and 
$\sum_{\{{\bf A}\}}$ defined as summing over $A_{ij}\in\{0,1\}$  for each $i<j$.

For any graph, we can define its degree matrix $\bf D$ which is diagonal with its diagonal entries $D_{ii}\equiv d_i$ as given  by (\ref{degree}), and the graph Laplacian matrix
\beq
{\bf L}={\bf D}-{\bf A}.
\label{Lapl}
\eeq
The graph Laplacian controls diffusion and random walks on graphs. Resistance distance $\Omega_{ij}$
between vertices $i$ and $j$ can be obtained as \cite{resdist5}
\begin{equation}
    \Omega_{ij} =L^{\mathrm{inv}}_{ii} + L^{\mathrm{inv}}_{jj} - 2L^{\mathrm{inv}}_{ij},
\label{Om}
\end{equation}
where ${\bf L}^{\mathrm{inv}}$ is the Moore-Penrose (pseudo)inverse of $\bf L$, defined so that ${\bf L}^{\mathrm{inv}}{\bf L}$
is the projector on the subspace spanned by all non-null eigenvectors of $\bf L$. If the graph is connected, $\bf L$ 
has only one null eigenvector proportional to $(1,1,\ldots,1)^T$ and hence ${\bf L}^{\mathrm{inv}}{\bf L}={\bf I}-{\bf 1}/N$, where $\bf I$ is the identity matrix and $\bf 1$ is the matrix all of whose entries equal 1.
We provide a derivation of (\ref{Om}) in Appendix~\ref{app:resdist}.

Our main object of study is the probability density for the resistance measured between vertices 1 and 2 to be equal $\rho$:
\begin{equation}
    P(\Omega_{12} = \rho) \equiv \langle \delta(L^{\mathrm{inv}}_{11}+L^{\mathrm{inv}}_{22}-2L^{\mathrm{inv}}_{12}- \rho)\rangle = \frac{1}{Z} \sum_{\{\mathbf{A}\}} \delta(L^{\mathrm{inv}}_{11}+L^{\mathrm{inv}}_{22}-2L^{\mathrm{inv}}_{12}- \rho)  \left(\frac\kk N\right)^{\sum_{i<j}A_{ij}}.
\label{Pdef}
\end{equation}
The corresponding distribution for any other pair of vertices would evidently be the same as the \ER ensemble enjoys complete vertex permutation symmetry. This function can be straightforwardly sampled numerically,
and such numerical experiments suggest that the distribution tends to a definite curve at large $N$ and displays rather sophisticated shapes. Our purpose is to develop some analytic theory for this distribution.

One can start by expressing the $\de$-function in (\ref{Pdef}) in terms of its Fourier representation:
\begin{equation}
    P(\rho) \propto \int_{-\infty}^\infty d\xi \sum_{\{\mathbf{A}\}} e^{i\xi(L^{\mathrm{inv}}_{11}+L^{\mathrm{inv}}_{22}-2L^{\mathrm{inv}}_{12}- \rho)}  \left(\frac\kk N\right)^{\sum_{i<j}A_{ij}}.
    \label{eq:prho}
\end{equation}
From this point on, we shall be ignoring the overall prefactor of $P(\rho)$ keeping in mind that it can always be recovered at the end by normalizing $P$ as a probability density,
\beq
\int_0^\infty d\rho \,P(\rho)=1.
\label{Pnorm}
\eeq

Summation over $\bf A$ still cannot be performed directly in (\ref{eq:prho}), but it will become possible after we introduce a set of auxiliary fields. We start with an $N$-dimensional real-valued vector $\pmb{\phi}$ whose components are $\phi_i$ and write,
using the standard multidimensional Gaussian integration, 
\begin{equation}
 e^{i\xi(L^{\mathrm{inv}}_{11}+L^{\mathrm{inv}}_{22}-2L^{\mathrm{inv}}_{12})}=  \left(\frac{i\xi}{\pi}\right)^{\frac{N-1}{2}}{\sqrt{\pdet{\mathbf{L}}}}\; \int d\pmb{\phi}\; \delta\left(\sumph\right) e^{-i\xi \sum_{kl} L_{kl}\phi_k \phi_l +2i\xi(\phi_1-\phi_2)}.
\label{eq:Hubbard}
\end{equation}
Such Gaussian inversion formulas are frequently employed in random matrix literature relying on statistical physics methods \cite{spectrum,MF,euclRM,ecological}; some brief pedagogical comments clarifying the structure of (\ref{eq:Hubbard}) are provided in Appendix~\ref{app:matinv}.
We have inserted $\delta(\sum_i \phi_i)$ inside the integral in (\ref{eq:Hubbard}) to control the null direction of $\mathbf{L}$ corresponding to the eigenvector $(1,1,\ldots,1)^T$, and $\det'{\bf L}$ denotes the pseudodeterminant of $\bf L$, that is the product of its nonzero eigenvalues. Then, using
$\sum_{ij}L_{ij}\phi_i\phi_j=\frac{1}{2}\sum_{ij}A_{ij}(\phi_i - \phi_j)^2= \sum_{i<j} A_{ij}(\phi_i - \phi_j)^2$,
we arrive at
\begin{equation}
    P(\rho) \propto \int d\xi \,\xi^\frac{N-1}{2} e^{-i\xi\rho}  \int d\pmb{\phi}\;\delta\left(\sumph\right) e^{2i\xi (\phi_1 - \phi_2)}\sum_{\{\mathbf{A}\}} \sqrt{\pdet{\mathbf{L}}}\,\, e^{\sum_{k<l}A_{kl}[\ln(\kk/N)-i\xi(\phi_k - \phi_l)^2]}.
\label{eq:poriginal}
\end{equation}

A few comments are in order before we proceed. First, (\ref{eq:Hubbard}) is only literally correct if the graph is connected and $(1,1,\ldots,1)^T$ is the only null eigenvector of its Laplacian, which is, strictly speaking, not true in our case. Nonetheless,
this inaccuracy will not impede our subsequent derivations in terms of $1/c$ expansions at $N\to\infty$, as we shall now explain. At $N\to\infty$, an \ER graph with a fixed $c>1$ consists almost surely \cite{newman} of a single giant connected component whose size is of order $N$ and further small components whose size is at most logarithmic in $N$. The probability $u$ for a given vertex not to belong to the giant component satisfies \cite{newman} the equation
\beq
u=e^{-c(1-u)}.
\eeq 
At large $c$, this implies $u\approx e^{-c}$. Therefore, the number of vertices outside the giant connected component of the \ER graph, and  certainly the number of disconnected components, and hence the number of distinct null eigenvectors of the Laplacian are all suppressed by $e^{-c}$. Thus, one may legitimately expect that their effect will be simply invisible within an $1/c$ expansion. Our derivations will indeed show no pathologies due to these neglected null eigenvectors, while the results will be in agreement with numerical simulations, which validates the intuition given above.

Second, the summation over $\bf A$ still cannot be performed directly in (\ref{eq:poriginal}) because of the presence of $\sqrt{\pdet{\mathbf{L}}}$. The determinant of the graph Laplacian is known to equal the number of spanning trees of the graph, and it has been studied for random graphs in \cite{span1,span2}. There are two possible approaches to handling that determinant, both of which we shall explore. At the most naive level, one may
expect some concentration behavior for this determinant at $N\to\infty$ and large $c$, which would turn it into an irrelevant numerical factor that can be taken outside the sum. The sum over $\bf A$ can then be immediately evaluated, leaving behind an integral over $\xi$ and $\pmb{\phi}$ accessible to statistical field theory methods. At a more refined level,
one may simply write an exact representation for $\sqrt{\pdet{\mathbf{L}}}$ by introducing a few further Gaussian integrals over extra auxiliary fields, a technique often employed in the physics of disordered systems \cite{efetov, fyodorov, FN}. After that, the summation over $\bf A$ can again be performed, but now without any approximations or guesses for $\sqrt{\pdet{\mathbf{L}}}$, and the result is still tractable by statistical field theory methods. As an outcome of implementing the more refined approach, one in fact observes that the naive approach of discarding the determinant had been completely justified at leading order in $1/c$, and receives explicit computable corrections at subleading orders, where the determinant should not be neglected.

Whether working with the naive approach, as we shall do in section~\ref{saddle}, or with the exact approach, as we shall do in section~\ref{sub}, evaluating the sum over $\bf A$ leaves behind an integral over variables defined on the graph vertices and interacting in a pair-wise manner. This integral may be viewed as a rather peculiar statistical field theory on a complete graph. Such field-theoretic representations have been successfully applied in the past
to a variety of random graph problems, as in \cite{PN}, see also \cite{spectrum,AC,corr}. The essence of these constructions is that the resulting field-theoretic integral admits a saddle-point evaluation, in our case, as an expansion in powers of $1/c$.

In view of our subsequent comparisons of the asymptotic evaluation of (\ref{Pdef}) with numerical simulations,
we conclude this section with a remark on how such comparisons are made. One can easily generate an adjacency matrix of an \ER graph and then invert it to obtain the resistance distance matrix according to (\ref{Om}).
It may be additionally wise to restrict the adjacency matrix to the giant component of the \ER graph, though it is not essential, at least at large $c$, since all the other components contain a small number of vertices and cannot significantly affect the statistics.
Once the resistance distance matrix has been obtained from (\ref{Om}), it would be wasteful to only use its $\Omega_{12}$ entry for constructing the distribution of the resistance distance between vertices 1 and 2 and discard the rest. Indeed, resistance distances between different pairs of vertices must be identically distributed due to the permutation symmetry of the \ER ensemble, and one also expects them to be rather weakly correlated since the entries of the adjacency matrix are completely uncorrelated. It is then natural to sample the whole set of values of $\Omega_{ij}$ corresponding to all pairs of vertices, and plot histograms of this sample. It is such samples that we use for testing our analytics,
and find convincing agreement, normally for $N=25000$ (this provides a giant number of entries, given by $N(N-1)/2$, for constructing the histograms).

\section{Saddle point analysis}\label{saddle}

We shall now proceed with a basic saddle point analysis of (\ref{eq:poriginal}), assuming that $\sqrt{\pdet{\mathbf{L}}}$ concentrates in an appropriate sense and can be treated as $\bf A$-independent within the \ER ensemble. An accurate justification will be given to this assumption in the next section. The saddle point analysis we present gives a concrete mathematical realization to the `mean field' principles common in statistical physics, and the idea that a large number of nearest neighbors given by $\kk$ simplifies the behavior of the system due to `averaging over neighbors' is very much in line with these principles.

\subsection{Summation over $\bf A$}\label{sumA}

Assuming that $\sqrt{\pdet{\mathbf{L}}}$ in (\ref{eq:poriginal}) is approximately constant, it can be taken outside the $\bf A$-sum and merged with the other normalization factors. Thereafter, the sum over $\bf A$ is immediately evaluated in a manner that closely parallels \cite{PN}, since the summand turns into a product of factors each of which depends only on a single entry of the matrix $\bf A$:
\begin{equation}
     \sum_{\{\mathbf{A}\}}\prod_{k<l} \; e^{A_{kl}[\ln(\kk/N)-i\xi(\phi_k - \phi_l)^2]} = \prod_{k<l}\left(1+\frac{\kk}Ne^{-i\xi(\phi_k - \phi_l)^2 }\right)= \exp\hspace{-1mm}\left[\sum_{k<l}\ln \left(1+\frac{\kk}Ne^{-i\xi(\phi_k - \phi_l)^2 }\right)\right].
\label{eq:intout}
\end{equation}
This formula follows from the tautological summation identity $\sum_{\alpha=0}^1e^{\alpha\beta}=1+e^\beta$.

At $N\to\infty$ and a fixed $c$, we can furthermore approximate the logarithm as 
\beq
\ln \left(1+\frac{\kk}Ne^{-i\xi(\phi_k - \phi_l)^2 }\right)\approx \frac{\kk}Ne^{-i\xi(\phi_k - \phi_l)^2}.
\label{logexp}
\eeq
Hence,  (\ref{eq:poriginal}) turns into
\begin{equation}
    P(\rho)\propto\int d\xi \,\xi^{\frac{N-1}{2}} e^{-i\xi\rho}\int d\pmb{\phi}\;\delta\left(\sumph\right)  e^H,
    \label{eq:prho_new}
\end{equation}
where $H$ is defined as
\begin{equation}
    H \equiv \frac{\kk}{N} \sum_{k<l}  e^{-i\xi(\phi_k- \phi_l)^2 } + 2i\xi(\phi_1-\phi_2)
    = \frac{\kk}{2N}\sum_{k\neq l} e^{-i\xi(\phi_k - \phi_l)^2 } + 2i\xi(\phi_1-\phi_2).
\label{eq:Ham}
\end{equation}
One aspect of (\ref{eq:prho_new}) that will play a crucial role hereafter is that $H$ appears in the exponent and features a term proportional to $c$. Hence, assuming that $c$ is large makes (\ref{eq:prho_new}) amenable to conventional saddle point techniques, resulting in a viable approach to evaluating $P(\rho)$ in terms of an $1/\kk$ expansion. As we shall see below, the leading term of this expansion does a convincing job at capturing the behavior of $P(\rho)$ even at values of $\kk$ that are not nominally `large,' such as $\kk=4$.

\subsection{The saddle point equation}\label{saddlep}

Given the apparent saddle point structure of (\ref{eq:prho_new}) at large $\kk$, we must identify the stationary points of $H$, given by solutions of\footnote{It may appear odd at the first sight that we are including the $\xi$-term in the saddle point equation, while it comes with no explicit dependence on the saddle point parameter $\kk$. It is important to keep in mind, however, that $\xi$ is an integration variable and all possible values enter the game. The relevant scaling of the resistance distance is $\rho\sim 1/c$ (one can immediately see this numerically, and it will also come out of our analysis) and the corresponding scaling of the Fourier-conjugate variable $\xi$ is $\sim c$, making the $\xi$-term in (\ref{eq:Ham}) relevant for the saddle point equation. The test of this approach to generating the saddle point expansion is that a well-formed asymptotic series in $1/\kk$ will be produced. Namely, if (\ref{eq:Ham}) is expanded in Taylor series around the saddle point defined as we have specified, the leading contribution will arise from the Gaussian part of (\ref{eq:prho_new}), and higher order terms in (\ref{eq:Ham}) will generate $1/\kk$ corrections, as evident from our treatment below and considerations of Appendix~\ref{app:sublead}. By contrast, if we had omitted the $\xi$-term from the saddle point equation, the saddle point solution would have been $\pmb{\phi}=0$, resulting in a trivial estimate of the resistance distance distribution $P(\rho)=\delta(\rho)$ that conveys no meaningful message beyond the smallness of $\rho$ at large $\kk$ and cannot be straightforwardly corrected in terms of a perturbative expansion in $1/\kk$.}
\begin{equation}
    0=\frac{\partial H}{\partial \phi_k} =-\frac{2i\kk\xi}{N}\sum_{l\neq k}\left[ e^{-i\xi(\phi_k-\phi_l)^2}(\phi_k - \phi_l)\right] + 2i\xi(\delta_{1,k}-\delta_{2,k}).
\label{eq:H_approx}
\end{equation}
Once $\pmb{\phi}_0$ that satisfies this equation has been found, it can be used to obtain a saddle point approximation of  (\ref{eq:prho_new}) by applying
\begin{equation}
\int d\pmb{\phi} \; \delta\left(\sumph\right) e^{H(\pmb{\phi})} \approx e^{H(\pmb{\phi}_0)}\int d\tilde{\pmb{\phi}}\; \delta\left({\textstyle\sum_i\tilde{\phi}_i}\right) e^{-\frac{1}{2}\sum_{ij}(-M_{ij})\tilde\phi_i\tilde\phi_j} = \frac{(-2\pi)^{\frac{N-1}{2}}}{\sqrt{\pdet{\mathbf{M}}}}\;e^{H(\pmb{\phi}_0)},
\label{eq:saddle_approx}
\end{equation}
where $\tilde{\pmb{\phi}}\equiv\pmb{\phi}-\pmb{\phi}_0$,  $\det'$ once again denotes the pseudodeterminant, and we have used $\sum_i \phi_{0i}=0$, as all configurations integrated over in (\ref{eq:prho_new}) satisfy this relation. The Hessian $\bf M$ is given by
\beq
M_{ij}=\frac{\del^2H}{\del\phi_i\del\phi_j}\Big|_{\pmb{\phi}=\pmb{\phi}_0}.
\label{Hess}
\eeq
Note that the $\delta$-function in (\ref{eq:saddle_approx}) simply controls the null direction $(1,1,\ldots,1)^T$ of the Hessian $\bf M$ (some further details on Gaussian integrals with null directions can be found in Appendix~\ref{app:matinv}).

We must then identify the solutions of (\ref{eq:H_approx}). It is natural to start by examining the saddle points that respect the symmetries of $H$. (An alternative is families of saddle points converted into each other by the symmetries. It is more exotic, though certainly not impossible, and the ultimate validation of our approach, here and in general, comes from comparisons with numerical simulations.)
As one can see  from \eqref{eq:Ham}, $H$ is invariant under arbitrary permutations of $\phi_3,\phi_4,\ldots,\phi_N$ 
as well as the symmetry $(\phi_1,\phi_2)\to(-\phi_2,-\phi_1)$. Since only configurations with $\sumph=0$ contribute to
(\ref{eq:prho_new}), the only saddle candidate that respects the symmetries of $H$ is
 $\pmb{\phi}_0 = (\phi_0,-\phi_0,0,0,\dots)$ with a yet-unknown $\phi_0$. For configurations of this form, (\ref{eq:H_approx}) is rewritten as
\begin{equation}
\begin{split}
    0 &= 
    \begin{cases}
     \dsty-\frac{2i\xi}{N}\kk e^{-4i\xi\phi_0^2}(2{\phi_0}) - 2i\xi\frac{N-2}{N}\kk e^{-i\xi\phi_0^2}\phi_0 + 2i\xi & \text{for}\; k = 1, \vspace{2mm}\\
     \dsty\frac{2i\xi}{N}\kk e^{-4i\xi\phi_0^2}(2{\phi_0}) + 2i\xi\frac{N-2}{N}\kk e^{-i\xi\phi_0^2}\phi_0 - 2i\xi & \text{for}\; k = 2, \vspace{2mm} \\
     \dsty\frac{2i\xi}{N}\left(\kk e^{-i\xi\phi_0^2}\phi_0 + \kk e^{-i\xi\phi_0^2}(-\phi_0)\right) & \text{for}\; k \ge 3.
    \end{cases}
    \label{eq:saddle}
\end{split}
\end{equation}
The last equation is identically satisfied, while the equations for $k=1$ and $k=2$ are equivalent to each other, and reduce at $N\to\infty$ to
\begin{equation}
    \kk\phi_0 = e^{i\xi\phi_0^2}.
    \label{eq:self_con}
\end{equation}

The saddle point equation (\ref{eq:self_con}) is closely related to the equation $W(x)e^{W(x)}=x$ defining the Lambert W function \cite{lambertW}. The solution can correspondingly be expressed through the Lambert W function as
\begin{equation}
\phi_0 (\xi)= \sqrt{\frac{W(-{2i\xi}/{\kk^2})}{-2i\xi}}.
\label{phiW}
\end{equation}
In practice, it may often be convenient to construct $\phi_0$ by solving the ODE
\begin{equation}
\frac{d\phi_0}{d\xi} = -\frac{\phi_0^3}{i+2\xi\phi_0^2}
\label{eq:ode}
\end{equation}
that can be readily derived from (\ref{eq:self_con}), with the initial condition $\phi_0(0)=1/\kk$.
We quote the series expansion for $\phi_0$ to give an impression of its behavior near the origin:
\begin{equation}
    \phi_0 (\xi)= \frac{1}{\kk} + \frac{i \xi}{\kk^3} - \frac{5 \xi^2}{2 \kk^5} - \frac{49 i \xi^3}{6 \kk^7} + \frac{243 \xi^4}{8 \kk^9}+\cdots.
\label{phi0exp}
\end{equation}
Evidently $\kk\phi_0$ only depends on $\xi/\kk^2$.

Finally, the value of $H$ at the saddle point is evaluated as
\begin{equation}
   H(\pmb{\phi}_0)= \frac{N\kk}{2}+ \frac{2}{\phi_0} + 4i\xi\phi_0\to \frac{2}{\phi_0} + 4i\xi\phi_0,
\label{H0}
\end{equation}
where the saddle point equation (\ref{eq:self_con}) has been used to simplify this expression, and the last `arrow' operation indicates that the $\xi$-independent term ${N\kk}/{2}$ can be merged into the normalization of $P(\rho)$ and hence ignored. What remains for completing the saddle point estimate (\ref{eq:saddle_approx}) is to evaluate the Hessian determinant $\pdet{\mathbf{M}}$.

\subsection{The Hessian determinant}

The second derivatives of $H$ can be expressed as
\begin{equation}
\frac{\partial^2 H}{\partial \phi_l \partial \phi_k} =
\begin{cases}
\dsty\frac{2i\xi\kk}{N}\sum_{i\neq k}  e^{-i\xi(\phi_k-\phi_i)^2}\left[2i\xi (\phi_k-\phi_i)^2-1\right] & \text{for}\; l = k,\vspace{2mm}\\
\dsty\frac{-2i\xi\kk}{N} e^{-i\xi(\phi_k-\phi_l)^2} \left[2i\xi (\phi_k-\phi_l)^2 - 1\right] &\text{for}\; l\neq k.
\end{cases}
\end{equation}
At the saddle point (\ref{eq:self_con}), this becomes
\begin{equation}
M_{lk} =\frac{\partial^2 H}{\partial \phi_l \partial \phi_k} \bigg\rvert_{\pmb{\phi}_0}=
\begin{cases}
\dsty -\frac{2\xi}{N}\left[\frac{1}{\phi_0^4\kk^3}(i+8\xi\phi_0^2)+(N-3)\frac{1}{\phi_0}(i+2\xi\phi_0^2)\right] \equiv a &\text{for}\; l=k\leq2,\vspace{1mm}\\
\dsty-\frac{2\xi}{N}\left[\frac{2}{\phi_0}(i+2\xi\phi_0^2)+(N-3)\kk\right] \equiv b &\text{for}\; l=k>2,\vspace{1mm}\\
\dsty\frac{2}{N\xi^2\phi_0^4}(i+8\xi\phi^2)\equiv g &\hspace{-4mm} \text{for}\; l=1,k=2,\vspace{1mm}\\
\dsty\frac{2\xi}{N\phi_0}(i+2\xi\phi^2)\equiv f & \hspace{-7.8mm}\text{for}\; l =1,2; k>2,\vspace{1mm}\\
\dsty \frac{2i\kk\xi}{N}\equiv d &\hspace{-8.3mm}\text{for}\; l\neq k\;, l,k > 2,
\end{cases}
\label{eq:Hess}
\end{equation}
where we have specified the values for $l\le k$, and the remaining values can be recovered using the symmetry of $M$.
The structure of this matrix can be visualized as follows:
\begin{equation}\mathbf{M} = 
\begin{pmatrix}
a & g & f & f & f & \dots \vspace{1mm}\\
g & a & f & f & f & \dots\vspace{1mm} \\
f & f & b & d & d & \dots \vspace{1mm}\\
f & f & d & b & d & \dots \\
f & f & d & d & b & \vdots \\
\vdots & \vdots & \vdots & \vdots & \dots & \ddots
\end{pmatrix}
\label{eq:Mmatrix}
\end{equation}
where $a, b,d,f$ and $g$ are defined in \eqref{eq:Hess}. Note that each row (and column) sums to $0$ since $H$ is not affected by a simultaneous common shift of all $\phi_k$. In other words,
\beq
a + g + (N-2) f = 0,\qquad
2f + b + (N-3)d = 0.
\label{eq:Hconstraint}
\eeq

The eigenvectors of $\bf M$ can be worked out explicitly. The vector $(1,1,\dots)^T$ is annihilated by $\bf M$, as follows from (\ref{eq:Hconstraint}). This null eigenvector and its zero eigenvalue will evidently not contribute to the computation of the pseudodeterminant $\pdet{\mathbf{M}}$.  The second obvious eigenvector is $(1,-1,0,0,\ldots)^T$ and its eigenvalue is $a-g$. Finally, there are $N-3$ eigenvectors of the form $(0,\ldots,0,1,-1,0,0\ldots)^T$ where the first ``1'' must occur in position 3 or higher. Their eigenvalues are $b-d$. We have explicitly recovered $N-1$ eigenvectors, and since $\bf M$ is a Hermitian $N\times N$ matrix, the unique remaining eigenvector must be orthogonal to all of them and hence proportional to  $(-1,-1,\frac{2}{N-2},\frac{2}{N-2},\frac{2}{N-2}\dots)^T$. Its eigenvalue can be computed by direct application of $\bf M$ and then simplified using \eqref{eq:Hconstraint} to give $-Nf$. With all the eigenvalues at hand, we can write down the desired pseudodeterminant explicitly as
\begin{equation}
\pdet \mathbf{M} = -Nf(a-g)(b-d)^{N-3}.
\label{eq:pdetM}
\end{equation}
Now, we can substitute \eqref{eq:Hess} back in (\ref{eq:pdetM}) and take the $N\to\infty$ limit. This requires some care in the last factor since $(1+{x}/{N})^N \rightarrow e^x$; elsewhere, we can simply discard all subleading contributions in $1/N$. Altogether,
\begin{equation}
\pdet\mathbf{M}\propto \xi^{N-1}\left(\frac{1}{\phi_0}-2i\xi\phi_0\right)^2e^{\frac{2}{\kk\phi_0}-\frac{4i\xi\phi_0}{\kk}},
\label{detM}
\end{equation}
where we have omitted all the irrelevant $\xi$-independent factors.

\subsection{The leading saddle point estimate}

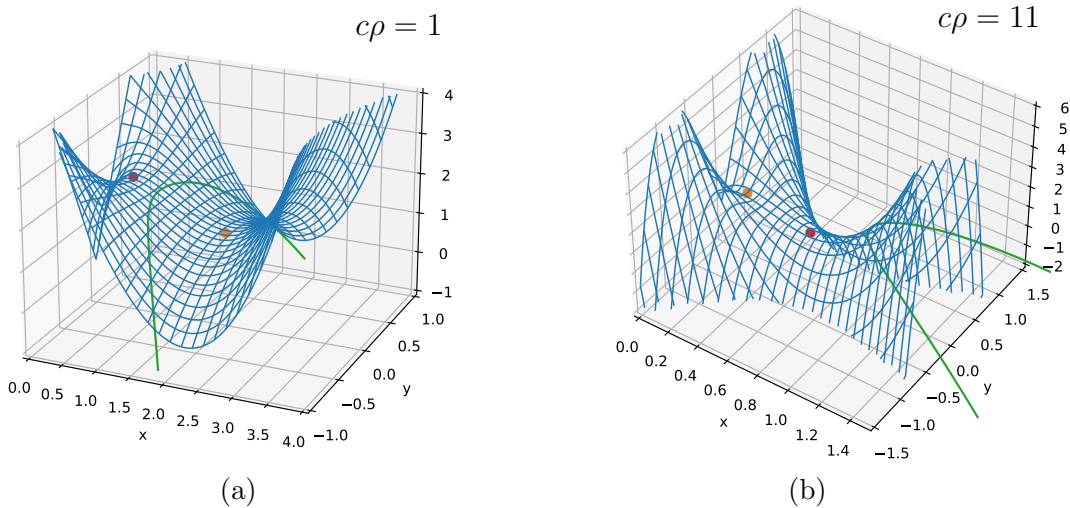
\begin{figure}[t]
\centering
\begin{subfigure}{0.45\textwidth}
\hspace{-1cm}\includegraphics[width = 1.2\linewidth]{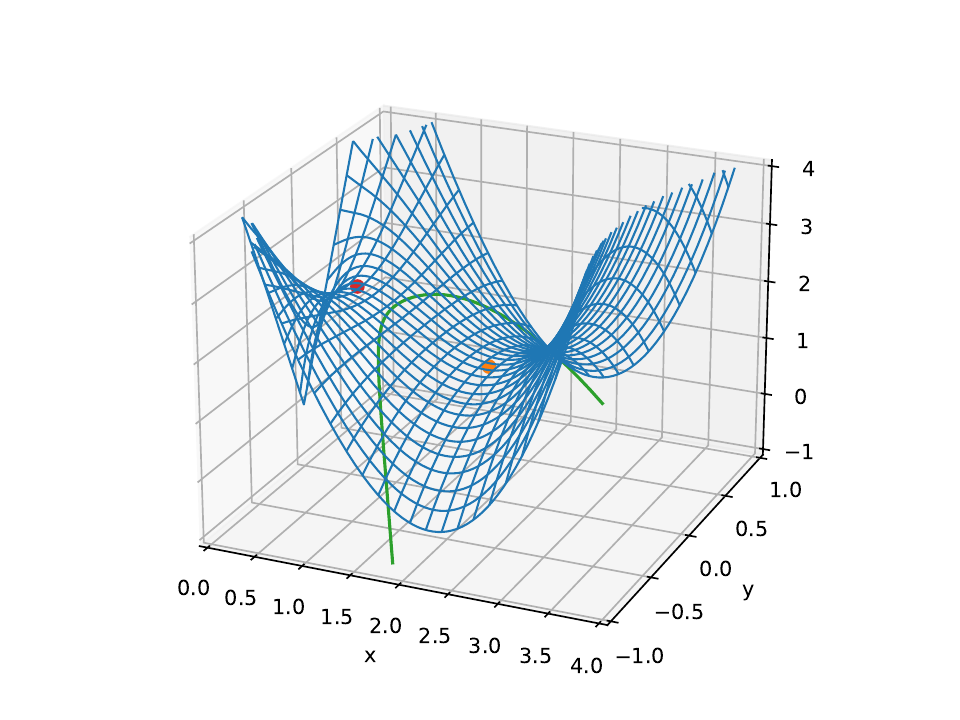}\vspace{-8mm}
\begin{picture}(0,0)
\put(150,160){$\kk\rho=1$}
\end{picture}
\caption{}
\end{subfigure}
\begin{subfigure}{0.45\textwidth}
\hspace{-5mm}\includegraphics[width = 1.2\linewidth]{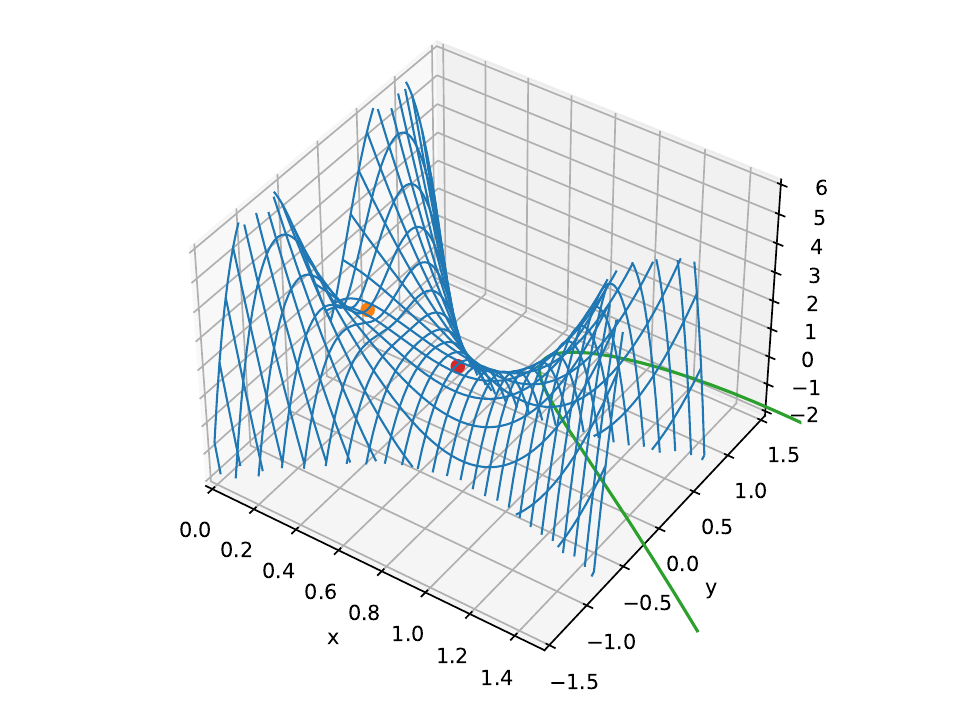}\vspace{-8mm}
\begin{picture}(0,0)
\put(155,163){$\kk\rho=11$}
\end{picture}
\caption{}
\end{subfigure}
\caption{The real part of the saddle point function $2z+(\rho\kk z^2-4z)\ln z$ of (\ref{eq:tp_endresult}), evaluated at $z=x+iy$. The orange dots represent the saddle at $z=2/\kk\rho$, and the red dots represent the saddle at $z=1/\sqrt{e}$. The green curves visualize the integration contour in (\ref{eq:tp_endresult}). The two plots demonstrate how the two saddle point locations change as $\kk\rho$ varies from $\kk\rho=1$ (a) to $\kk\rho = 11$ (b). }
\label{fig:saddles}
\end{figure}
Putting together (\ref{eq:prho_new}), (\ref{eq:saddle_approx}), (\ref{H0}) and (\ref{detM}), we obtain the leading order saddle point evaluation of (\ref{Pdef}) in the form
\begin{equation}
P(\rho)\propto  \int d\xi \,\,\left(\frac{1}{\phi_0}-2i\xi\phi_0\right)^{-1} e^{-i\xi\rho} e^{\frac{2}{\phi_0}\left(1-\frac{1}{2\kk}\right)+4i\xi\phi_0\left(1+\frac1{2c}\right)},
\label{eq:FTcorr1}
\end{equation}
with $\phi_0$ defined by (\ref{eq:self_con}) and conveniently reconstructed by solving (\ref{eq:ode}). Normalization (\ref{Pnorm}) should be applied.
This expression for $P(\rho)$ is the main result of this paper. We shall now recast it in a few alternative forms, and provide further simplified formulas valid at large $\kk$.

A useful observation is that the pre-exponential factor in (\ref{eq:FTcorr1}) can be expressed in terms of $d\phi_0/d\xi$ using (\ref{eq:ode}), which gives
\begin{align}\label{eq: changevar}
	P(\rho) &\propto \int_{-\infty}^{\infty} d \xi \,\frac{d \phi_0}{d\xi}\,\frac{1}{\phi_0^2}\,e^{\frac{2}{\phi_{0}}\left(1-\frac{1}{2\kk}\right)+4i\xi\phi_0\left(1+\frac1{2\kk}\right)}.
\end{align}
As a result, one can choose $\phi_0$ to be the new integration variable, which is convenient since, while $\phi_0(\xi)$ is expressed through the Lambert W function, its inverse $\xi(\phi_0)= -i\ln\left(\kk \phi_0\right)/\phi_0^2$ is an elementary function. A particularly convenient choice for the integration variable is $z=1/(\kk\phi_0)$, which gives
\beq
	P(\rho) \propto \int_{1/\kk\phi_0(-\infty)}^{1/\kk\phi_0(\infty)} dz \ e^{2\kk z\left(1-\frac{1}{2\kk}\right)}z^{ \rho \kk^2z^2-\left( 4\kk + 2\right)z}.
\label{eq:tp_endresult}
\eeq
The integration contour, as inherited from (\ref{eq: changevar}), comes from infinity within the first quadrant $(\Re z>0,\Im z>0)$, passes through $z=1$ and exists to infinity via the fourth quadrant $(\Re z>0,\Im z<0)$. The contour can evidently be deformed freely, as with any complex plane integral, as long as it does not touch the singular branching point at $z=0$.

As the integrand features $\kk$ in the exponent, and we are treating $\kk$ as a large parameter, it is natural to apply further saddle point evaluation to (\ref{eq:tp_endresult}). To this end, we note that the logarithm of the integrand is
$-2z\ln z -z + \kk[2z+(\rho\kk z^2-4z)\ln z]$. Demanding that the leading part of this expression (given in the square brackets) is stationary at $z_0$ yields the saddle point equation for the $z$-integral:
\beq
(\kk\rho z_0-2)(1+2\ln z_0) = 0.
\eeq
This equation evidently has two solutions, $z_0 = {2}/\kk\rho$ and $z_0= 1/\sqrt{e}$. The topography
of the integrand is such that the contour can always be deformed to pass through $\max({2}/{\kk\rho}, 1/\sqrt{e})$,
while the second saddle lies between the contour and the singularity at $z=0$ and is thus irrelevant for the saddle point evaluation. The location of the contour relatively to the saddles is depicted in Fig.~\ref{fig:saddles}. Extracting the leading saddle point estimate due to the appropriate saddles thus yields
\begin{figure}[t]
\centering
\begin{subfigure}{0.45\textwidth}
\hspace{-1.2cm}\includegraphics[width=1.2\textwidth]{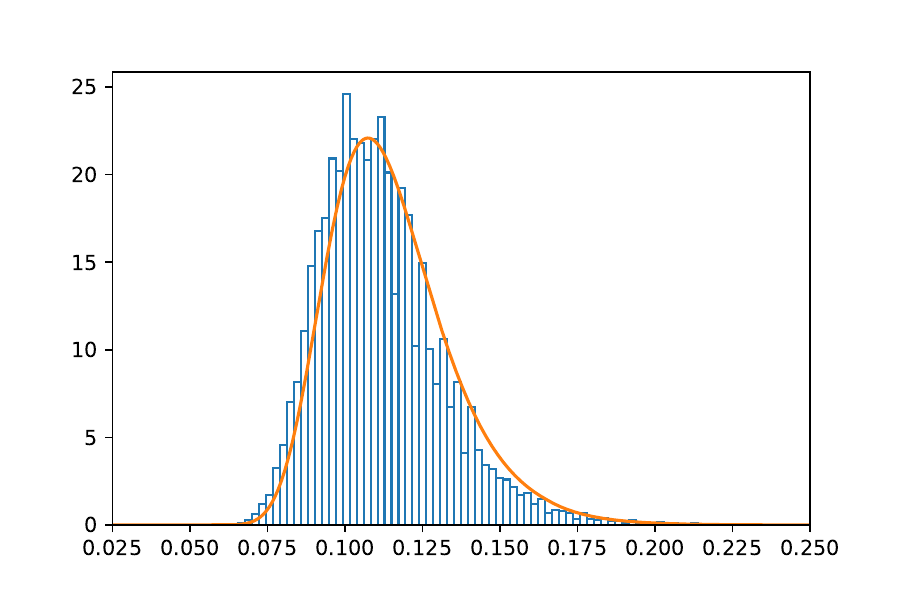}\vspace{-8mm}
\begin{picture}(0,0)
\put(1,127){$P$}
\put(182,18){$\rho$}
\put(155,127){$\kk=20$}
\end{picture}
\caption{\rule{1.3cm}{0mm}}
\label{fig:c20}
\end{subfigure}
\begin{subfigure}{0.45\textwidth}
\includegraphics[width = 1.2\textwidth]{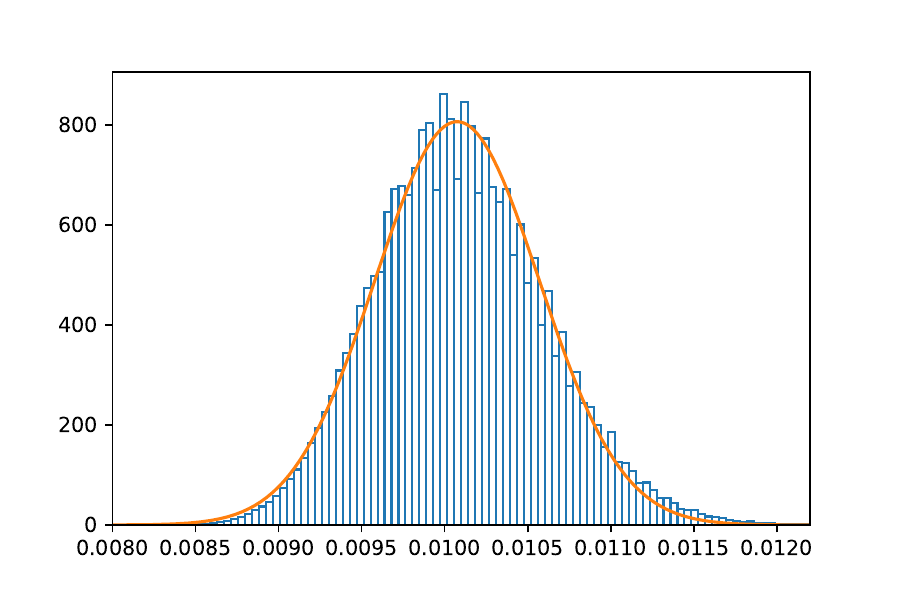}\vspace{-8mm}
\begin{picture}(0,0)
\end{picture}
\begin{picture}(0,0)
\put(31,127){$P$}
\put(214,18){$\rho$}
\put(181,127){$\kk=200$}
\end{picture}
\caption{\rule{-1.8cm}{0mm}}
\label{fig:c200}
\end{subfigure}
\caption{Resistance distance distribution of the largest connected component of an \ER graph with 25000 vertices at mean degree 20 (a) and 200 (b). Numerical simulation results are plotted as blue histograms, while the saddle point estimate (\ref{skew}) for (a) and the large $\kk$ Gaussian approximation (\ref{eq:Gapprox}) for (b) are plotted as orange curves.}
\label{fig:clarge}
\end{figure}
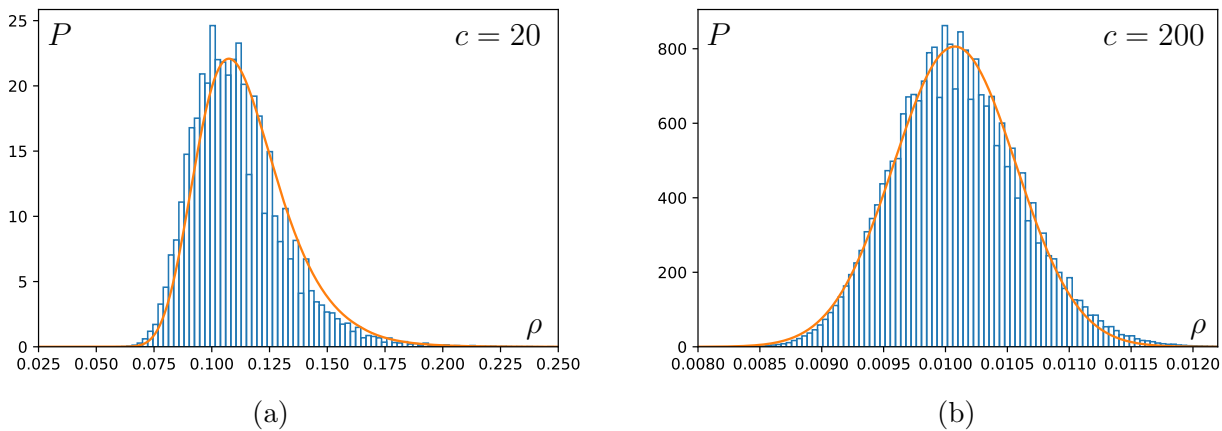
\beq
P(\rho) \propto
\begin{cases}
\dsty \left(\frac{\kk\rho}{2}\right)^{\frac{4}{\kk\rho}}e^{-\frac{2}{\kk\rho}}\left(\frac{e\kk\rho}{2}\right)^{\frac{4}{\rho}} & \text{for}\; \rho\le2\sqrt{e}/\kk,\vspace{1mm}\\
\exp\left(\frac{4\kk}{\sqrt{e}} - \frac{\rho\kk^2}{2e}\right) &\text{for} \; \rho>2\sqrt{e}/\kk.
\end{cases}
\label{skew}
\end{equation}
We have not attempted recovering the pre-exponential factors in this saddle point estimate, which is slightly subtle due to the coalescing saddles \cite{wong} at $\rho=2\sqrt{e}/\kk$, and would not have improved the accuracy significantly in the regime of interest.
We compare this curve to the actual numerical simulations at the moderate value $\kk=20$ in Fig.~\ref{fig:c20}, showing a good agreement.

Finally, when $\kk$ is very large, it makes sense to expand (\ref{skew}) around its maximum at $\rho\approx 2/\kk$ which yields a Gaussian approximation of the form
\beq
P(\rho) \propto \exp\left[-\frac{\kk^3}{4\left(1-{4}/{\kk}\right)}\left(\rho-\frac{2}{\kk}-\frac{3}{\kk ^2}\right)^{\hspace{-1mm}2\hspace{1mm}}\right].
\label{eq:Gapprox}
\eeq
This distribution is again in an excellent agreement with the numerics at large $c$, see Fig.~\ref{fig:c200}. Asymptotically, the distribution tends to a Gaussian of mean $2/\kk$ and standard deviation $\sqrt{2/\kk^3}$.

\begin{figure}[t]
\centering
\begin{subfigure}{0.45\textwidth}
\hspace{-1.2cm}\includegraphics[width=1.2\textwidth]{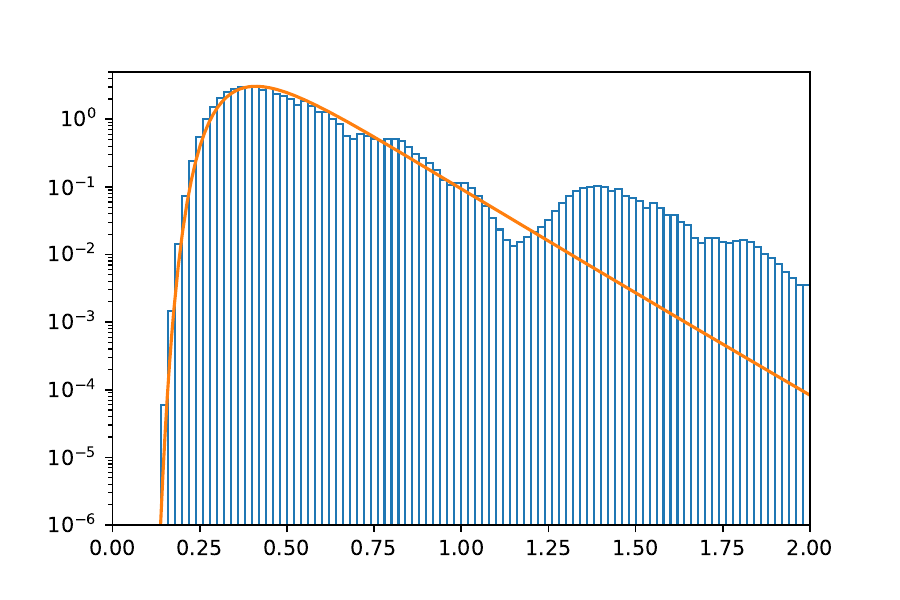}\vspace{-8mm}
\begin{picture}(0,0)
\put(1,127){$P$}
\put(182,18){$\rho$}
\put(162,127){$\kk=6$}
\end{picture}
\caption{\rule{1.3cm}{0mm}}
\label{fig:c6}
\end{subfigure}
\begin{subfigure}{0.45\textwidth}
\includegraphics[width = 1.2\textwidth]{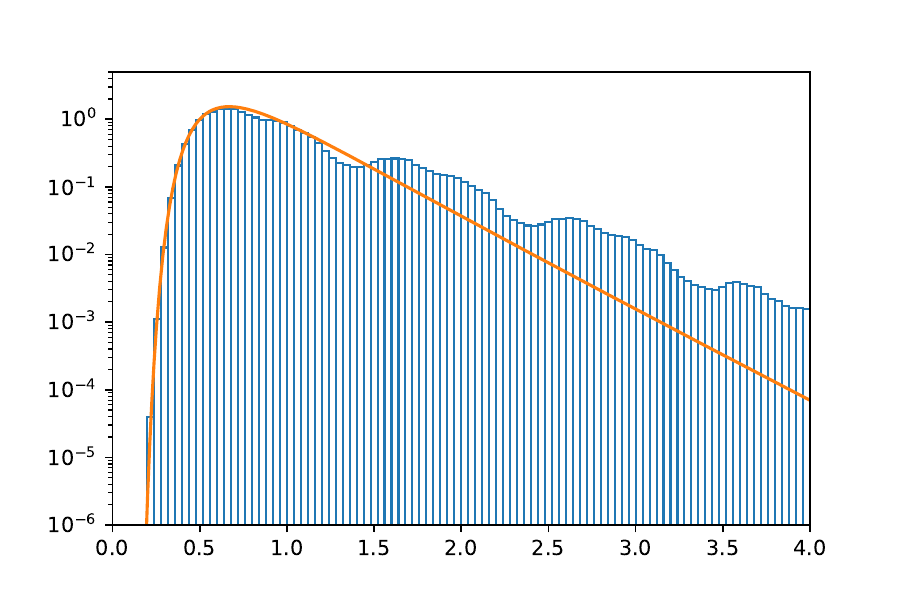}\vspace{-8mm}
\begin{picture}(0,0)
\end{picture}
\begin{picture}(0,0)
\put(31,127){$P$}
\put(212,18){$\rho$}
\put(191,127){$\kk=4$}
\end{picture}
\caption{\rule{-1.8cm}{0mm}}
\label{fig:c4}
\end{subfigure}
\caption{Resistance distance distribution of the largest connected component of an \ER graph with 25000 vertices at mean degree 6 (a) and 4 (b). Numerical simulation results are plotted as blue histograms, while the saddle point estimate (\ref{eq:FTcorr1}) is plotted as orange curves. We have used the log-linear scale to emphasize that, while the main peak is captured adequately, there is excess probability in the right tail compared to our saddle point estimate. We shall revisit the contributions at larger values of $\rho$ in section~\ref{deg1}.}
\label{fig:c86}
\end{figure}
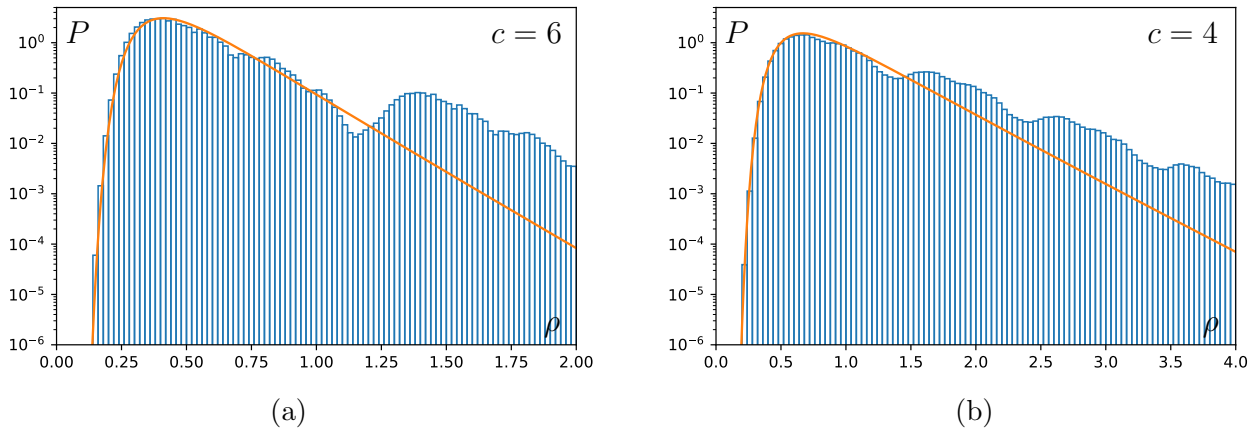
The agreement between our saddle point estimate and numerical simulations continues to hold even for such low values of $\kk$ as 6 or 4. In this case, (\ref{eq:FTcorr1}) gives a slightly better match than (\ref{skew}), and that is what we plot in Fig.~\ref{fig:c86}. To quantify the agreement of the analytic and empirical distributions, we use the {\it overlapping coefficient} \cite{overlap}, which is defined, for any two normalized probability distributions $p_1(x)$ and $p_2(x)$, as $\int  \min(p_1(x),p_2(x))\,dx$. By definition, the overlapping coefficient is a number between 0 and 1 that equals 1 if and only if the two distributions are identical. For our analytic curve as compared to the numerics, this gives approximately 92\% for $\kk=6$ and approximately 90\% for $\kk=4$. We emphasize that the saddle point estimate (\ref{eq:FTcorr1}) only captures the main peak of the distribution without reproducing accurately its right tail. We shall return to the excess probability observed to the right of the main peak in section~\ref{deg1}.

\section{Fluctuations of $\sqrt{\pdet L}$ in (\ref{eq:poriginal})}\label{sub}

This section can be comfortably skipped by practically-minded readers that are interested in obtaining useful approximations to the resistance distance distribution, rather than in justifying the corresponding derivations. Our purpose here is to close an essential gap in the reasoning of the previous section, though after this has been accomplished, the result of the previous section will remain unchanged.

We would like to return to (\ref{eq:poriginal}) and provide an accurate treatment of the $\sqrt{\pdet L}$ factor that has been neglected in the previous section. This is done by introducing a few more auxiliary fields and produces a more elaborate structure similar to (\ref{eq:prho_new}-\ref{eq:Ham}). As a result of these considerations, we shall see that (a) the analysis of the previous section is justified at leading order in $1/\kk$, (b) the determinant does contribute to subleading corrections in $1/\kk$ (though such contributions are not crucial for the practical purposes of this paper). More precisely, when the determinant contribution and other higher order corrections are taken into account, after an appropriate `linked cluster' resummation (the technical details are given in Appendix~\ref{app:sublead}), the exponent of the last factor in (\ref{eq:FTcorr1}) turns into a series in powers of $1/\kk$. However, the powers of $1/\kk$ already written in (\ref{eq:FTcorr1}) remain unchanged by these contributions, which only supply higher (and hence more strongly suppressed) powers of $1/\kk$.

To proceed with our derivations, we shall represent the determinant factor $\sqrt{\pdet L}$ in terms of Gaussian integrals involving both ordinary (commuting) and Grassmanian (anticommuting) variables. This approach has often been taken up in statistical physics considerations related to disordered systems \cite{efetov,fyodorov,FN}. To implement this approach, we rewrite
\begin{equation}
\sqrt{\pdet L} = \frac{1}{\sqrt{\pdet L}}\pdet{L}.
\label{detsplit}
\end{equation}
The first factor can be represented as an ordinary Gaussian integral
\beq
\frac{1}{\sqrt{\pdet L}} = \pi^{-\frac{N-1}{2}}\int d\pmb\chi\,\delta\left({\textstyle\sum_i \chi_i}\right) e^{-\sum_{ij}L_{ij}\chi_i\chi_j}.
\label{det1}
\eeq
For the second factor, a convenient representation is given in terms of anticommuting (Grassmanian) fields \cite{efetov} denoted $\pmb{\theta}$ and $\pmb{\eta}$ that satisfy
\beq
\theta_i\theta_j = -\theta_j\theta_i \qquad(\text{so that} \quad \theta_i^2 = 0),\qquad
\int d\theta_i\;\theta_j  = \de_{ij}, \qquad \int d\theta_i = 0,
\eeq
and similar relations for the components of $\pmb{\eta}$, while the componenents of $\pmb{\theta}$ and $\pmb{\eta}$ anti-commute with each other. With the {\it Berezin integration rules} given above (a more detailed explanation can be found in \cite{efetov}), we express the determinant $\pdet{L}$ in terms of these new fields as
\beq
\pdet L = \int d\pmb\theta d\pmb\eta\;\,\delta\left({\textstyle\sum_i \theta_i}\right)\,\delta\left({\textstyle\sum_i \eta_i}\right) e^{-\sum_{ij}L_{ij}\theta_i\eta_j}.
\label{det2}
\eeq
Observing that $\sum_{ij}L_{ij}\chi_i\chi_j = \sum_{i<j} A_{ij}(\chi_i-\chi_j)^2$ and $L_{ij}\theta_i\eta_j =\sum_{i<j}A_{ij}(\theta_i-\theta_j)(\eta_i-\eta_j)$, and inserting (\ref{detsplit}) represented using (\ref{det1}) and (\ref{det2}) into (\ref{eq:poriginal}), one can perform the summation over $\bf A$ in a manner identical to (\ref{eq:intout}). We then arrive
at the following representation of (\ref{eq:poriginal}), with the determinant factor now fully taken into account:
\begin{equation}
P(\rho)\propto \int d\xi \,\xi^{(N-1)/{2}} e^{-i\xi\rho}  \int d\pmb{\phi}d\pmb\chi d\pmb\theta d\pmb\eta\;\delta\left(\sumph\right)\,\delta\left({\textstyle\sum_i \chi_i}\right)\,\delta\left({\textstyle\sum_i \theta_i}\right)\,\delta\left({\textstyle\sum_i \eta_i}\right)e^{H'}.
\label{Pfull}
\end{equation}
Here, $H'$ is defined by
\beq
H' = \frac{\kk}{N}\sum_{i<j}\exp\left[-i\xi(\phi_i-\phi_j)^2-(\chi_i-\chi_j)^2-(\theta_i-\theta_j)(\eta_i-\eta_j)\right]+2i\xi(\phi_1-\phi_2).
\label{Hprime}
\eeq
We have applied the $N\to\infty$ limit, as in (\ref{logexp}), to simplify the above expression.

The new representation (\ref{Pfull}-\ref{Hprime}) can be treated by saddle point methods in direct parallel
to our previous treatment of (\ref{eq:prho_new}-\ref{eq:Ham}). The saddle point configuration for $\pmb\phi$ is 
exactly the same as in section~\ref{saddlep} and is still defined by the saddle point equation (\ref{eq:self_con}).
All the remaining fields in (\ref{Hprime}) vanish in the saddle point configuration. As a result, the different fields involved completely decouple from each other in the expansion of (\ref{Hprime}) up to quadratic order in the fields about the saddle point: 
\beq
H' =H(\pmb{\phi}_0)+\frac12\sum_{ij}M_{ij}\tilde\phi_i\tilde\phi_j-\frac{c}{N}\sum_{i<j}[(\chi_i-\chi_j)^2+(\theta_i-\theta_j)(\eta_i-\eta_j)],
\eeq
where the first two terms are defined by (\ref{H0}) and (\ref{eq:Hess}). Since the quadratic forms of
$\pmb\chi$, $\pmb\theta$ and $\pmb\eta$ in the above expression decouple from the quadratic form of $\tilde{\pmb\phi}$, and are furthermore $\xi$-independent,
the leading order saddle point estimate remains completely identical to section~\ref{saddle}. Thus, our neglect of the $\sqrt{\pdet{L}}$ factor in the derivations of section~\ref{saddle} is {\it a posteriori} justified. 

At the same time, the different fields involved no longer decouple if one expands (\ref{Hprime}) beyond the quadratic order in the fields, and thus extra corrections at higher orders in $1/\kk$ will be produced by the additional fields in (\ref{Hprime}) that capture the fluctuations of $\sqrt{\pdet{L}}$. We provide an analysis of these corrections in Appendix~\ref{app:sublead}, though they do not play a significant practical role in the considerations of this paper.

\section{Subleading corrections and subleading peaks}\label{deg1}

One can systematically compute the higher order corrections to (\ref{eq:FTcorr1}) within the $1/\kk$ expansion using the language of Feynman diagrams. We give a technical demonstration of such computations at first subleading order in Appendix~\ref{app:sublead}. In practice, however, these corrections are numerically insignificant at large $\kk$, while at small $\kk$, as in Fig.~\ref{fig:c86}, they do not improve the agreement with the numerics. In fact, they slightly shift the peak to the right away from the excellent match in Fig.~\ref{fig:c86}, without inducing any other significant changes. A skeptic might then say that the agreement we observe when the leading order estimate is extended to small values of $\kk$ is coincidental. We, however, feel that it must be systematic as it is observed for different small values of $\kk$ and there must be an underlying mathematical reason for it, beyond the scope of our current comprehension. Be it as it may, at this stage, we simply report the agreement of the leading order saddle point estimate with the numerical data at low $\kk$ as an empirical fact.

\begin{figure}
\centering
\begin{subfigure}{0.45\textwidth}
\hspace{-1.2cm}\includegraphics[width=1.2\textwidth]{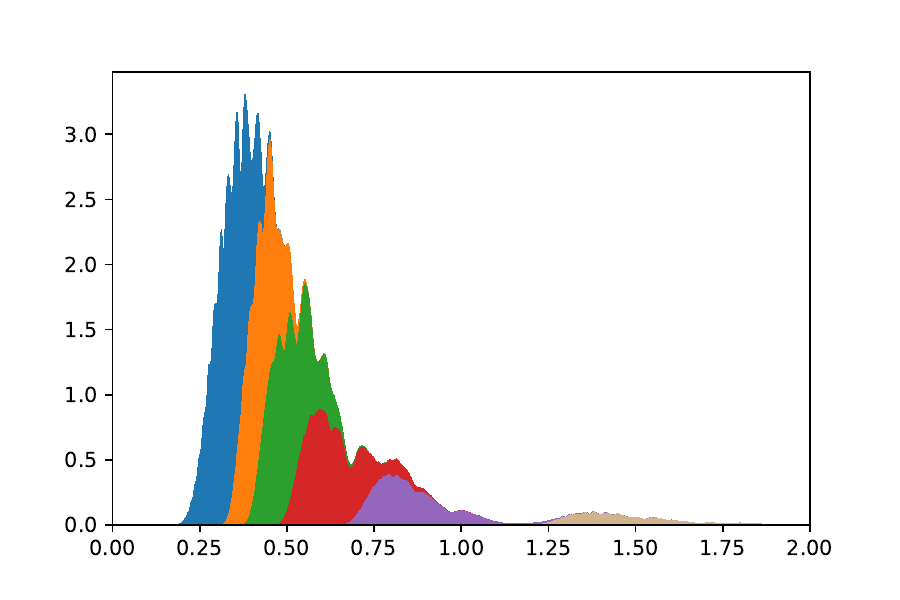}\vspace{-8mm}
\begin{picture}(0,0)
\put(1,127){$P$}
\put(182,18){$\rho$}
\end{picture}
\caption{\rule{1.3cm}{0mm}}
\label{fig:color}
\end{subfigure}  
\begin{subfigure}{0.45\textwidth}
\includegraphics[width = 1.2\textwidth]{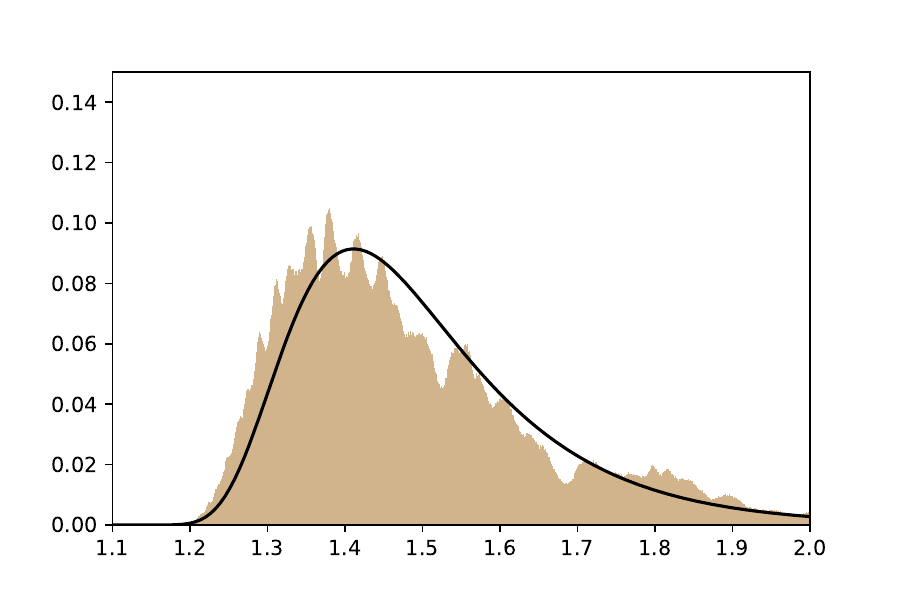}\vspace{-8mm}
\begin{picture}(0,0)
\put(37,127){$2P_1$}
\put(216,22){$\rho$}
\end{picture}
\caption{\rule{-1.8cm}{0mm}}
\label{fig:deg1}
\end{subfigure}
\caption{Refined representations for the resistance distance distribution of the largest connected component of an \ER graph with 25000 vertices at mean degree 6: (a) the contributions of the different pairs of vertices to the resistance distance  histogram obtained from numerical simulations, colored according to the smaller value of the vertex degrees within each pair; the colors, appearing in sequence from right to left, correspond to the smaller degree values of 1, 2, 3, 4, 5, and all the remaining values lumped together; (b) the (rightmost) contribution to the histogram in (a) involving vertices of degree 1, isolated and compared to the analytic prediction $2P_1(\rho)$ extracted as described under (\ref{P1res}) and plotted as a black curve.}
\end{figure}
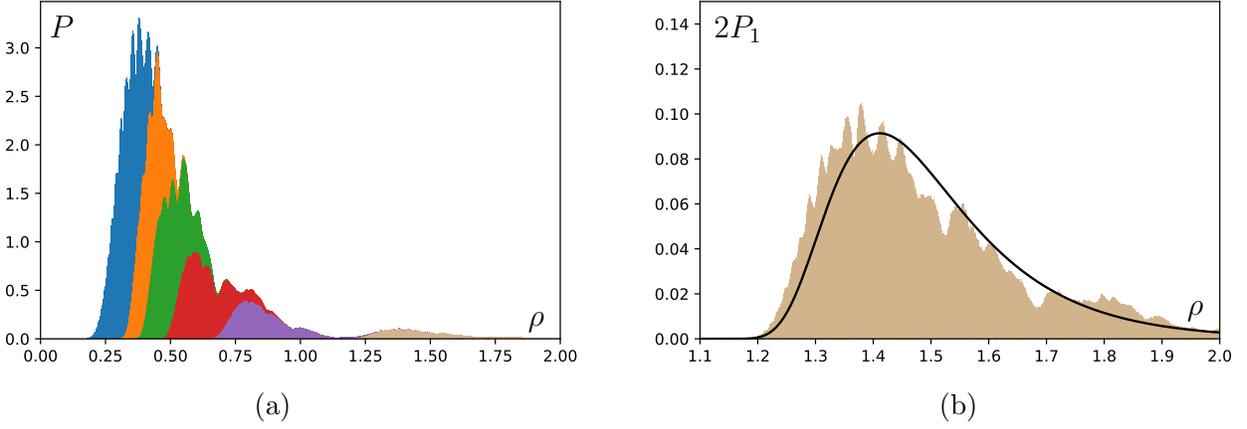
Figs.~\ref{fig:clarge} and \ref{fig:c86} show that the leading order saddle point estimate does a great job at reproducing the main outline of the resistance distance probability distribution $P(\rho)$. Closer examination, however, reveals more detailed features of the empirical probability distributions, in particular, at low values of $\kk$.
To demonstrate this aspect, we have given in Fig.~\ref{fig:color} a more detailed version of the empirical resistance distance histogram at $\kk=6$ that we have previously displayed in a more coarse-grained presentation in Fig.~\ref{fig:c6}, this time switching to a linear rather than log-linear plot.
(We ask the reader to ignore the colors in Fig.~\ref{fig:color} at this point.)
First, there are sharp narrow sub-peaks that decorate the main peak and, in fact, may suggest a lack of smoothness in the true distribution. We do not have a theoretical understanding of these sharp peaks and will briefly comment on them further in the conclusions. Second, there are `bumps' that are seen on top of the tail of the distribution that extends to higher values of $\rho$. In the rest of this section, we aim to improve our understanding of these `bumps.'

To demonstrate the nature of the subleading peaks at large $\rho$ to the right of the main peak, we have  made a further refinement in Fig.~\ref{fig:color}. Namely, when plotting the empirical histogram of resistance distances $\Omega_{ij}$ between vertices $i$ and $j$ in a computer-generated \ER graph, we have colored the contributions according to the minimal degree of the two vertices involved, $\min(d_i,d_j)$. This coloring reveals that the rightmost visible peak around $\rho\in(1.25,1.75)$ in fact entirely consists of resistance distances between vertices of degree 1 and the rest of the graph. For other peaks at lower values of $\rho$, the picture is more complicated, but it is still clear that these peaks are made from contributions involving vertices of low degrees. (Qualitatively similar peaks due to vertices of specific low degrees have been seen in the analysis of a different characteristic of random graphs in \cite{hetero}.)

The above empirical observations suggest that it is natural to approach the analysis of the subleading peaks
by introducing degree-differentiated resistance distance distributions that not only specify the resistance distance between vertices 1 and 2, but also require that the degree of vertex 1  equals $d$, namely
\begin{equation}
    P_d(\Omega_{12} = \rho) = \frac{1}{Z} \sum_{\{\mathbf{A}\}} \delta\left(L^{\mathrm{inv}}_{11}+L^{\mathrm{inv}}_{22}-2L^{\mathrm{inv}}_{12}- \rho\right)\delta\left(d,\textstyle{\sum_j} A_{1j}\right) \; \left(\frac{\kk}N\right)^{\sum_{i<j}A_{ij}}.
\label{degdiffP}
\end{equation}
Evidently,
\beq
\sum_{d=1}^\infty P_d(\rho)=P(\rho),
\label{Pdecomp}
\eeq
thus one can think of $P_d$ as controlling the different contributions to $P(\rho)$ in a more discerning manner.
By writing a Fourier representation for the $\de$-function involving $\rho$, as in (\ref{eq:prho}), and a complex plane representation for the Kronecker symbol involving $d$,
\beq
\delta\left(d,\textstyle{\sum_j} A_{1j}\right)=\frac1{2\pi i} \oint\frac{dz}{z^{d+1}} \prod_j z^{ A_{1j}},
\eeq
one obtains
\beq
P_d(\rho) \propto \oint\frac{dz}{z^{d+1}} \int d\xi\sum_{\{\mathbf{A}\}} e^{i\xi(L^{\mathrm{inv}}_{11}+L^{\mathrm{inv}}_{22}-2L^{\mathrm{inv}}_{12} - \rho)}\prod_j z^{ A_{1j}} \prod_{i<j}\left(\frac{\kk}N\right)^{A_{ij}}.
\eeq
Then, as in section~\ref{setup}, we can write
\begin{align}
P_d(\rho) \propto& \oint\frac{dz}{z^{d+1}} \int d\xi\, e^{-i\xi\rho} \xi^{\frac{N-1}2} \int d\pmb{\phi}\,\, \delta\left(\sumph\right) e^{2i\xi(\phi_1-\phi_2)}\\
&\hspace{2cm}\times\sum_{\{\mathbf{A}\}} \sqrt{\pdet{\mathbf{L}}}\, e^{-i\xi\sum_{k<l}A_{kl}(\phi_k-\phi_l)^2}\prod_j z^{ A_{1j}} \prod_{i<j}\left(\frac{\kk}N\right)^{A_{ij}}.\nonumber
\end{align}
We could have given an accurate treatment of the $\sqrt{\pdet{\mathbf{L}}}$ factor, as in section~\ref{sub}, by introducing extra auxiliary fields, but just like in section~\ref{sub}, the result (at leading order in $1/\kk$) would be exactly the same as what one gets by treating $\sqrt{\pdet{\mathbf{L}}}$ as constant. We shall therefore simply omit this factor to keep the formulas more compact, without affecting the result. Then, performing the summation over $\bf A$ as in section~\ref{sumA}, we get
\begin{align}
P_d(\rho) \propto& \oint\frac{dz}{z^{d+1}} \int d\xi\, e^{-i\xi\rho} \xi^{\frac{N-1}2} \int d\pmb{\phi}\,\, \delta\left(\sumph\right) e^{2i\xi(\phi_1-\phi_2)}\label{degdiffPz}\\
&\hspace{2cm}\times\prod_{j\ge 2}\left(1+\frac{cz}{N}e^{ -i\xi(\phi_1-\phi_j)^2}\right)\prod_{2\le k<l}\left(1+\frac{c}{N}e^{ -i\xi(\phi_k-\phi_l)^2}\right).\nonumber
\end{align}

We now focus on $d=1$. The contour integral featuring $dz/z^2$ simply extracts the first $z$-derivative of the integrand at $z=0$. This produces a factor of $\frac{\kk}{N}\sum_{j\ge 2} e^{-i\xi(\phi_1-\phi_j)^2}$. The contribution from $j=2$ can be ignored as it is suppressed by $1/N$ and vanishes at $N\to\infty$. The remaining $N-2$ contributions from $j\ge 3$ will all equal each other upon integration since all the other factors in the integrand are invariant under all permutations of $(\phi_3,\phi_4,\ldots,\phi_N)$. As a result, 
$\frac{\kk}{N}\sum_{j\ge 2} e^{-i\xi(\phi_1-\phi_j)^2}$ can be effectively replaced by $\kk\,e^{-i\xi(\phi_1-\phi_3)^2}$ at $N\to\infty$, leaving
\beq
P_1(\rho) \propto \int d\xi\, e^{-i\xi\rho} \xi^{\frac{N-1}2} \int d\pmb{\phi}\,\, \delta\left(\sumph\right) e^{2i\xi(\phi_1-\phi_2)-i\xi(\phi_1-\phi_3)^2}
\prod_{2\le k<l}\left(1+\frac{\kk}{N}e^{ -i\xi(\phi_k-\phi_l)^2}\right).
\eeq
We now introduce an $(N-1)$-dimensional vector $\pmb{\phi}'$ whose components are $\phi'_j=\phi_{j+1}+\phi_1/(N-1)$ so that $\sum_{j=1}^N\phi_j=\sum_{j=1}^{N-1}\phi'_j$. Then,
\begin{align}
P_1(\rho) \propto &\int d\xi\, e^{-i\xi\rho} \xi^{\frac{N-1}2} \int d\pmb{\phi}'\,\, \delta\left({\textstyle\sum_i}\phi'_i\right) \exp\left[\sum_{ k<l}\ln{\textstyle\left(1+\frac{\kk}{N}e^{ -i\xi(\phi'_k-\phi'_l)^2}\right)}\right]\nonumber\\
&\times\int d\phi_1 e^{-2i\xi[\phi'_1-N\phi_1/(N-1)]}e^{-i\xi[\phi'_2-N\phi_1/(N-1)]^2}.
\end{align}
The integral over $\phi_1$ is Gaussian and can be straightforwardly evaluated to yield a result proportional to $e^{i\xi}e^{-2i\xi(\phi'_1-\phi'_2)}/\sqrt{\xi}$, so that
\beq
P_1(\rho) \propto \int \hspace{-1mm}d\xi\, e^{-i\xi(\rho-1)} \xi^{\frac{N-2}2} \int \hspace{-1mm}d\pmb{\phi}'\,\, \delta\left({\textstyle\sum_i}\phi'_i\right) \exp\left[\sum_{ k<l}\ln{\textstyle\left(1+\frac{\kk}{N}e^{ -i\xi(\phi'_k-\phi'_l)^2}\right)}-2i\xi(\phi'_1-\phi'_2)\right].
\eeq
But after interchanging $\phi'_1$ and $\phi'_2$ one notices that, with $c\ll N$, this is the same expression as (\ref{eq:prho_new}-\ref{eq:Ham}), except that $\rho$ got replaced by $\rho-1$ and $N$ got replaced by $N-1$. Hence, in the $N\to\infty$ limit,
\beq
P_1(\rho)\propto P(\rho-1).
\eeq
To identify the numerical coefficient in this relation, we notice that while $P(\rho)$ is normalized to 1 by (\ref{Pnorm}), the integral of $P_1(\rho)$ over $\rho$ is simply the probability for vertex 1 to have degree 1, as one can see by directly integrating the definition (\ref{degdiffP}). The degree distribution of sparse \ER graphs is known to be Poissonian \cite{newman} and hence the probability for vertex 1 to have degree 1 is $\kk e^{-\kk}$. Hence,
\beq
P_1(\rho)=\kk e^{-\kk}P(\rho-1).
\label{P1res}
\eeq

In Fig.~\ref{fig:deg1}, we have plotted (\ref{P1res}) against the distribution of resistance distances involving vertices of degree 1 from a numerical simulation of an \ER graph at $\kk=6$. The saddle point estimate (\ref{eq:FTcorr1}), shifted to the right by one unit has been used to approximate $P(\rho-1)$ on the right-hand side of (\ref{P1res}). To get the normalization right, one must remember that in $P_1$ defined according to (\ref{degdiffP}) the degree of vertex 1 is required to be 1, while in our histogram, either of the two vertices involved in a particular resistance distance value can have degree 1. For that reason, one should compare the histogram to $2P_1(\rho)$. With these specifications, we find convincing agreement between the two curves (remembering the rather low value of $\kk$ and the asymptotic nature of our saddle point approach). 

The result (\ref{P1res}) is rather intuitive in the following sense. To reach the rest of the graph from a degree 1 vertex, the electric current must traverse the edge that connects this vertex to its unique nearest neighbor, which adds 1 ohm to the resistance distance, and then reach the target vertex starting from this unique nearest neighbor. It is natural to imagine that the resistance distance contributed by this second stage is distributed in the same way as the resistance distance between two generic vertices. This intuitive picture suggests that the distribution of resistance distances involving vertices of degree 1 should mimic the general distribution of resistance distances, but is shifted by 1 ohm, as in (\ref{P1res}). Our analysis provides a mathematical foundation for this intuitive picture.

It is evident due to the  suppression by $e^{-\kk}$ that $P_1(\rho)$, which contributes to $P(\rho)$ according to (\ref{Pdecomp}), could not possibly be recovered by computing $1/\kk$ corrections to the saddle point estimate for $P(\rho)$ developed in section~\ref{saddle}. It remains an open question whether subleading saddles exist in (\ref{eq:prho_new}-\ref{eq:Ham}) that would make it possible to recover the `bump' in $P(\rho)$ arising from $P_1(\rho)$ directly from (\ref{eq:prho_new}-\ref{eq:Ham}),
without resorting to the decomposition (\ref{Pdecomp}).

It should be possible to repeat our derivations for $P_d$ with $d\ge 2$. The construction will still involve applying the Cauchy formula to (\ref{degdiffPz}) and then evaluating a Gaussian integral over $\phi_1$. However, the source term in the resulting formula (the term linear in $\pmb\phi$ appearing in the exponent) will no longer exactly match (\ref{eq:prho_new}-\ref{eq:Ham}). So one can still attempt a saddle point analysis at large $\kk$ as in section~\ref{saddle}, but the technical details will change. Intuitively, this complication corresponds to the fact that, at higher $d$, the current can exit the source vertex toward any of its $d$ nearest neighbors and then has to reach the target vertex from any of these neighbors, which is a more complicated picture than for $d=1$. We believe it is very likely that an analytic representation can be developed for $P_d$ with $d\ge 2$, but prefer to limit ourselves here to the relatively straightforward case $d=1$ treated above.

\section{Discussion}

We have proposed an auxiliary field representation in the spirit of statistical field theory for the resistance
distance distribution in large \ER graphs of fixed mean degree $\kk$. Using this representation, a saddle point estimation of this distribution becomes possible at large $\kk$ in terms of the saddle point equation (\ref{eq:self_con}), producing the analytic curve (\ref{eq:FTcorr1}), which at progressively larger values of $\kk$ can be further simplified to (\ref{skew}) or even to the Gaussian form (\ref{eq:Gapprox}). We have furthermore identified the subleading peaks observed at large resistance distances in numerical simulations with contributions of vertices of low degrees, and developed an analytic estimate (\ref{P1res}) for the rightmost prominent peak of this sort, coming from vertices of degree 1.

A few asymptotic approximations have been employed in our analysis, and it may be wise to summarize them here. First, starting from (\ref{logexp}), we only retained the leading term in $c/N$, thus focusing on the sparse graph regime $c\ll N$. We have furthermore resorted to an $1/c$ expansion in our analysis of 
(\ref{eq:prho_new}-\ref{eq:Ham}), making our derivations geared toward the asymptotic regime $1\ll c\ll N$, where they are validated by Fig.~\ref{fig:clarge}. It is a practical observation of Fig.~\ref{fig:c86} that our asymptotic analysis qualitatively captures the main peak of the distribution even for smaller values of $\kk$, strictly speaking, outside the intended validity domain of our approximations.

Our derivations have been heuristic in nature, in the sense that our plausible (but not rigorously justified) assumptions about the saddle point that dominates the quantity of interest have produced results that are convincing in terms of how they compare to the corresponding numerical simulations, even for low values of $\kk$ such as 6 or 4. One may hope that further mathematical work will provide error bounds for our approximate results, though this would likely require methods beyond what we have used to derive the formulas presented here. Even at a more basic level, within the context of the effective field-theoretic description (\ref{eq:prho_new}-\ref{eq:Ham}) or (\ref{Pfull}-\ref{Hprime}), one could hope to obtain a more thorough picture of the full set of saddle points and the explicit integration contour deformation that leads to our estimate, in particular in relation to the multiple branches of the Lambert W function in (\ref{phiW}). Our derivation has simply assumed that the relevant saddle point respects the symmetries of the integrand, and that one must use the main branch \cite{lambertW} of the Lambert W function, and the test of this assumption has been in successful comparisons of its outcome with numerical experiments.

Another relevant subtlety is that we have essentially treated the \ER graphs as if they were connected,
which gives a negligible mistake in an $1/\kk$ treatment at large $\kk$ since the fraction of vertices outside the giant connected component is suppressed as $e^{-\kk}$ \cite{newman}. However, the issue becomes more and more pressing as one decreases $\kk$. In particular, if one is interested in the critical ($\kk\approx 1$) or subcritical ($\kk<1$) regime of the \ER graphs (as in related considerations of \cite{short3} for shortest path distances), it is certain that the issue can no longer be ignored. It may still be possible to develop an auxiliary field representation along the lines of section~\ref{setup}, but a technical obstruction is that a Gaussian representation is needed for the pseudoinverse of a general matrix, without any prior assumptions about its null eigenvectors.

It must be possible, with relative ease, to adapt our considerations to more sophisticated random graph models that can themselves be effectively treated by statistical field theory methods. One natural starting point is the two-star model \cite{PN,AC} and its relatives \cite{corr}. Note that, for the \ER case, the averaging over the graphs is very simple and could be performed directly, but we had to employ a statistical field representation for the resistance distance $\de$-function necessary to obtain the corresponding probability distribution. For more complicated random graph ensembles, statistical field theory methods may be necessary to effectively handle the average over the graphs as well.

We conclude with some tentative comments on the intriguing small-scale features seen in the numerical simulations, such as the sharp narrow sub-peaks visible in Fig.~\ref{fig:color}. These features are not statistical noise (as one might imagine at the first sight) and neither go away with averaging over a large number of runs nor change their positions between different runs. Our analytic results say nothing about these sharp sub-peaks, though they do a great job at capturing the shape of the large-scale peak upon which these small features are superposed. One natural thought is that, since resistance distance values $1/d_i+1/d_j$ in terms of the vertex degrees $d_i$ and $d_j$ play a significant role in many considerations of resistance distances \cite{lovasz,lostspace1}, there could be sharp peaks in the distribution around the rational values $1/n+1/m$ coming from vertices of degrees $n$ and $m$. We have not,  however, been able to straightforwardly trace the locations of sharp sub-peaks in Fig.~\ref{fig:color} back to such simple rational numbers. Be it as it may, the ornate structure of Fig.~\ref{fig:color}, which is also replicated in the degree-differentiated resistance distance distributions presented in color in the same figure, makes it very tempting to conjecture that the true resistance distance distribution of an infinitely large \ER graph is not, in fact, a smooth curve. Refuting or supporting this conjecture would evidently require methodology considerably beyond the scope of this article.

\section*{Acknowledgments}

We thank Eytan Katzav for inciting our interest in properties of resistance distances, and for comments on the manuscript. Research of O.E. has been supported by the CUniverse project (CUAASC) at Chulalongkorn University. 
Research of T.C. has been supported by the Program Management Unit for Human Resources and Institutional Development, Research and Innovation (grant number B05F630108), and by Thailand Science Research and Innovation Fund Chulalongkorn University [CU\_FRB65\_ind (5)\_110\_23\_40].

\appendix

\section{Pedagogical account of the resistance distance}
\label{app:resdist}

Consider a graph defined by an adjacency matrix $\bf A$. Imagine that each edge is made of a conducting wire with resistance 1 ohm, and a current of 1 amp is injected into vertex $i$ and ejected from vertex $j$. This process induces voltage readings $u_k$ as measured at vertex $k$. Since 1 amp of current flows between vertices $i$ and $j$, the resistance measured between these two vertices, which is by definition the resistance distance, is 
\beq
\Omega_{ij}=u_j-u_i.
\label{Omu}
\eeq 
Our purpose in this appendix is to derive the expression (\ref{Om}) for this quantity.

The total current flowing out of vertex $k$ is 1 if $k=i$, $-1$ if $k=j$ and 0 otherwise. Let $l$ be a nearest neighbor of $k$ (so that $A_{kl}=1$). Since the edge between $k$ and $l$ is a wire of resistance 1 ohm, this edge is traversed by the current $u_l-u_k$. Summing this expression over all the edges originating from vertex $k$, we get the total current exiting vertex $k$, that is
\beq
\sum_{l: A_{kl}=1} (u_l-u_k)\equiv \sum_{l=1}^N A_{kl}(u_l-u_k)=x_k,
\eeq
where
\beq
x_k\equiv
    \begin{cases}
      \phantom{-}1 & \text{if}\ k=i, \\
     -1 & \text{if}\ k=j, \\
      \phantom{-}0 & \text{otherwise}.
    \end{cases}
\eeq
Since $\sum_l A_{kl}=d_k$ by (\ref{degree}), we can rewrite this formula in matrix notation as
\beq
{\bf Lu}= -{\bf x},
\label{Laplsys}
\eeq
where $\bf L$ is the Laplacian (\ref{Lapl}), $\bf u$ is the vector with components $u_k$, and $\bf x$ is a vector with components $x_k$.

Assume first that the graph described by $\bf A$ is connected. In that case, $\bf L$ has one null eigenvector $(1,1,\ldots,1)^T$. This vector represents a common shift of the voltages $u_k$ by a constant, which evidently does not affect the physics. Because of this null vector, $\bf L$ is not invertible. We can, however, modify it so as to be invertible by introducing the matrix $\bf 1$ all of whose entries equal 1, and replacing (\ref{Laplsys}) by
\beq
({\bf L}+\lambda{\bf 1}){\bf u}={\bf x},
\label{Laplsysshift}
\eeq
where $\lambda$ is an arbitrary real number. Since ${\bf L1}={\bf 1L}=0$ and ${\bf 1x}=0$, the above equation is equivalent to (\ref{Laplsys}) together with the extra condition ${\bf 1u}=0$, which is the same as $\sum_k u_k=0$, thus fixing the voltage shift ambiguity. Then,  (\ref{Laplsysshift}) is immediately solved by 
\beq
{\bf u}=-({\bf L}+\lambda{\bf 1})^{-1}{\bf x}, 
\eeq
where $({\bf L}+\lambda{\bf 1})^{-1}$ is the ordinary matrix inverse of ${\bf L}+\lambda{\bf 1}$. Equivalently, in components,
\beq
u_k=-({\bf L}+\lambda{\bf 1})^{-1}_{ki}+({\bf L}+\lambda{\bf 1})^{-1}_{kj},
\eeq
which is also a solution of (\ref{Laplsys}), for any $\lambda$.
Finally, from (\ref{Omu}),
\beq
\Omega_{ij}=({\bf L}+\lambda{\bf 1})^{-1}_{ii}+({\bf L}+\lambda{\bf 1})^{-1}_{jj}
-2({\bf L}+\lambda{\bf 1})^{-1}_{ij}.
\label{LinvOm}
\eeq
As emphasized already, the inverses are defined for any nonzero $\lambda$ and the above expression cannot depend on $\lambda$ due to the relation ${\bf L1}={\bf 1L}=0$. In particular, taking $\lambda$ to $\infty$ converts $({\bf L}+\lambda{\bf 1})^{-1}$ to the Moore-Penrose pseudoinverse of $\bf L$, yielding (\ref{Om}) as desired. Note that, for practical computations, (\ref{LinvOm}) with an arbitrarily specified $\lambda$ is often more convenient \cite{resdist5} than the Moore-Penrose pseudoinverse.

If the graph corresponding to $\bf A$ is disconnected, it is a matter of convention which value of resistance distance to assign to two vertices that belong to two different connected components, since no current may flow between two such vertices. Furthermore, $\bf A$ and $\bf L$ are block-diagonal with different blocks corresponding to different connected components. In this situation, there are more than one null vector (one null vector for each connected component) and the ordinary matrix inverses used in (\ref{LinvOm}) no longer exist. Nonetheless, the definition (\ref{Om}) based on the Moore-Penrose pseudoinverse still works, and it assigns resistance distance zero to any two vertices belonging to two different connected components, while resistance distances within the same connected component are the same as what one would get with (\ref{LinvOm}) by treating this particular component as an isolated connected graph. Thus, (\ref{Om}) provides a convenient prescription for defining resistance distances in disconnected graphs as well, which is commonly adopted in the literature.

\section{Gaussian integration formulas for matrix inversion}
\label{app:matinv}

We start with the well-known Gaussian integration formula
\begin{equation}
 e^{i\sum_{kl}M^{-1}_{kl} x_k x_l}=  \left(\frac{i}{\pi}\right)^{{N}/{2}}{\sqrt{\det{\mathbf{M}}}}\; \int d\pmb{\phi}\;  e^{-i \sum_{kl} M_{kl}\phi_k \phi_l +2i\sum_k \phi_k x_k},
\label{Gaussinv}
\end{equation}
and would like to use it to justify (\ref{eq:Hubbard}). An easy way to verify (\ref{Gaussinv}) is by introducing the orthonormal eigenbasis of $\bf M$ denoted ${\bf e}_k$ and the corresponding eigenvalues $\mu_k$. After changing the integration variables from $\pmb{\phi}$ to $y_k\equiv (\pmb{\phi},{\bf e}_k)$ (an orthogonal change of variables with a unit Jacobian), the $N$ integrations decouple into one-dimensional Gaussian (more precisely, Fresnel) integrals, and then (\ref{Gaussinv}) is recovered due to $\det{\mathbf{M}}=\prod_k\mu_k$ and $M^{-1}_{kl}=\sum_n (e_n)_k (e_n)_l/\mu_n$.

We cannot apply (\ref{Gaussinv}) for cases when $\bf M$ is proportional to the graph Laplacian $\bf L$ defined by (\ref{Lapl}), since in that case $\bf M$ possesses a null eigenvector ${\bf e}_1=(1,1,\ldots,1)^T$ with $\mu_1=0$. As a result, neither is the ordinary matrix inverse well-defined, nor is the Gaussian integral on the right-hand side of (\ref{Gaussinv}) convergent. This complication can be remedied, however, by considering the following modified formula:
\begin{equation}
 e^{i\sum_{kl}M^{\mathrm{inv}}_{kl} x_k x_l}=  \left(\frac{i}{\pi}\right)^{\frac{N-1}{2}}{\sqrt{\pdet{\mathbf{M}}}}\; \int d\pmb{\phi}\;  \delta\left(\sumph\right)e^{-i \sum_{kl} M_{kl}\phi_k \phi_l +2i\sum_k \phi_k x_k},
\label{Gaussinvnull}
\end{equation}
where $\pdet{\mathbf{M}}=\prod_{k=2}^N \mu_k$ is the pseudodeterminant and $M^{\mathrm{inv}}_{kl}=\sum_{n=2}^N (e_n)_k (e_n)_l/\mu_n$ is the pseudoinverse. Indeed, changing to the eigenbasis 
$y_k\equiv (\pmb{\phi},{\bf e}_k)$ as before, one obtains $\int dy_1 \delta(y_1) e^{2iy_1  (\pmb{x},{\bf e}_1)}=1$ for the $y_1$-integration, while the remaining integrations work in exactly the same way as in the derivation of (\ref{Gaussinv}).

Finally, substituting ${\bf M}=\xi {\bf L}$ and ${\bf x}=(\xi,-\xi,0,\ldots,0)^T$ into (\ref{Gaussinvnull}), we arrive at (\ref{eq:Hubbard}).

\section{Subleading corrections in the $1/\kk$ expansion}
\label{app:sublead}

We present here computations for the $1/\kk$ corrections to the leading order saddle point estimate (\ref{eq:FTcorr1}). While this material is of only peripheral importance for the main claim of the paper that the leading saddle point estimate provides a good approximation to the empirical data, it may be useful for future work to have these subleading computations collected here.

To analyze the $1/\kk$ corrections, we introduce $\varphi_1=\phi_1- \phi_0$ and $ \varphi_2=\phi_2-\phi_0$ and expand (\ref{Hprime}) around the saddle point configuration $\pmb{\phi}_0 = (\phi_0,-\phi_0,0,0,\dots)$, $\pmb{\chi}=\pmb{\theta}=\pmb{\eta}=0$ up to quartic order in the deviations, with $\tilde{\pmb{\phi}}\equiv \pmb{\phi}-\pmb{\phi}_0=(\varphi_1,\varphi_2,\phi_3,\ldots,\phi_N)$:
\begin{align}
H' &=H(\pmb{\phi}_0)+\frac12\sum_{ij}M_{ij}\tilde\phi_i\tilde\phi_j-\frac{c}{2N}\sum_{ij}[(\chi_i\hspace{-0.7mm}-\hspace{-0.7mm}\chi_j)^2+(\theta_i\hspace{-0.7mm}-\hspace{-0.7mm}\theta_j)(\eta_i\hspace{-0.7mm}-\hspace{-0.7mm}\eta_j)]+\Delta H+\Delta H', \label{eq:Hprime}\\
\Delta H&=-\frac{\kk\,\xi^2}{2N}\sum_{3\le i<j}(\phi_i-\phi_j)^4\label{eq:Hexpand}\\
&\phantom{=}+\frac{\xi^2}{N\phi_0}\sum_{j\ge3}\left( - \frac{2}{3}\phi_0(3-2i\xi\phi_0^2)(\varphi_1-\phi_j)^3+\frac{1}{6}(-3+12i\xi\phi_0^2+4\xi^2\phi_0^4)(\varphi_1-\phi_j)^4\right)\nonumber\\
&\phantom{=}+\frac{\xi^2}{N\phi_0}\sum_{j\ge3}\left(\frac{2}{3}\phi_0(3-2i\xi\phi_0^2)(\varphi_2-\phi_j)^3 +\frac{1}{6}(-3+12i\xi\phi_0^2+4\xi^2\phi_0^4)(\varphi_2-\phi_j)^4\right),\nonumber\\
\Delta H'&=\frac{i\xi}{N}\frac{1-2i\xi\phi_0^2}{\phi_0}\left\{\sum_{j\ge 3} (\varphi_1-\phi_j)^2[(\chi_1-\chi_j)^2+(\theta_1-\theta_j)(\eta_1-\eta_j)]\right.\label{DeltaHprimeexpand}\\
&\hspace{4cm}\left.+\sum_{j\ge 3} (\varphi_2-\phi_j)^2[(\chi_2-\chi_j)^2+(\theta_2-\theta_j)(\eta_2-\eta_j)]\right\}+\nonumber\\
&\hspace{-1cm}+\frac{i\xi\kk}{N}\sum_{3\le i<j}(\phi_i-\phi_j)^2[(\chi_i\hspace{-0.7mm}-\hspace{-0.7mm}\chi_j)^2+(\theta_i\hspace{-0.7mm}-\hspace{-0.7mm}\theta_j)(\eta_i\hspace{-0.7mm}-\hspace{-0.7mm}\eta_j)]+\frac{1}{2} \sum_{i<j}\left((\chi_i\hspace{-0.7mm}-\hspace{-0.7mm}\chi_j)^4 -(\theta_i\hspace{-0.7mm}-\hspace{-0.7mm}\theta_j)^2(\eta_i\hspace{-0.7mm}-\hspace{-0.7mm}\eta_j)^2\right).\nonumber
\end{align}
Here, $H(\pmb{\phi}_0)$ is given by (\ref{H0}), $M_{ij}$ is given by (\ref{eq:Hess}), and the saddle point equation (\ref{eq:self_con}) has been used to eliminate the terms linear in $\tilde{\pmb{\phi}}$ and to simplify the other terms. To organize the  expressions above, we have separated the summation over $k$ and $l$ in (\ref{Hprime}) into four regions, which we demonstrate as follows for the case of of (\ref{eq:Hexpand}):
\begin{itemize}
\item $k=1,l=2$ -- this term can be discarded at $N\to\infty$ as it is explicitly suppressed by $1/N$; it is not included in (\ref{eq:Hexpand}),
\item $k=1$, $l\ge3$ -- these terms give rise to the second line in (\ref{eq:Hexpand}),
\item $k=2$, $l\ge3$ -- these terms give rise to the third line in (\ref{eq:Hexpand}),
\item $k,l\ge3$ -- these terms give rise to the first line in (\ref{eq:Hexpand}).
\end{itemize}
A similar treatment has been applied in (\ref{DeltaHprimeexpand}).
One should keep in mind that $\bf M$ is of order $\kk$, and hence ${\bf M}^{-1}$ is of order $1/\kk$, and $\phi_0$ is of order $1/\kk$ according to (\ref{phi0exp}).
In all the computations, two kind of terms should be systematically ignored: the terms that vanish at $N\to\infty$ and the terms that are $\xi$-independent (the latter can only contribute to the irrelevant normalization factor of $P(\rho)$ as will become apparent below). For example, the last sum in the last line of 
 (\ref{DeltaHprimeexpand}) can only produce $\xi$-independent contributions in the first order corrections.

In (\ref{eq:Hexpand}), one can see terms cubic in $\tilde{\pmb{\phi}}$. These terms, however, come with an extra suppression in $1/\kk$ and cannot possibly contribute to the leading order corrections. Indeed, one can convince oneself that all even order terms in (\ref{eq:Hexpand}) come with coefficients $O(\kk)$, while all odd order terms are $O(1)$; it should be kept in mind that $\phi_0$ is $O(1/\kk)$. This suppression is easily understood because (\ref{eq:Hexpand}) arises from expanding a function even under reflections of $\pmb{\phi}$ around a small value of $\pmb{\phi}$ defined by $\phi_0$. As a result, contributions of the cubic terms are suppressed and irrelevant at leading order. The first nonvanishing correction comes from the square of the cubic terms in (\ref{eq:Hexpand}), and it can be seen to contribute at order $1/\kk^3$.
There are similar cubic terms in (\ref{DeltaHprimeexpand}), but they similarly cannot contribute and have already been omitted in (\ref{DeltaHprimeexpand}) in order not to clutter the formulas.

With all of these preliminaries, the relevant corrections can be defined as $\mathcal{F}_{4,\phi}$ and ${\cal F}_{4,\chi\theta\eta}$ given by
\begin{align}
 \label{eq:saddle_det}&\int d\pmb{\phi}d\pmb\chi d\pmb\theta d\pmb\eta\;\delta\left(\sumph\right)\,\delta\left({\textstyle\sum_i \chi_i}\right)\,\delta\left({\textstyle\sum_i \theta_i}\right)\,\delta\left({\textstyle\sum_i \eta_i}\right)\\
&\times e^{\frac12\sum M_{ij}\tilde\phi_i\tilde\phi_j-\frac{c}{2N}\sum[(\chi_i-\chi_j)^2+(\theta_i-\theta_j)(\eta_i-\eta_j)]}(1+\Delta H+\Delta H')\propto\frac{(-2\pi)^{\frac{N-1}{2}}}{\sqrt{\pdet{\mathbf{M}}}}\;\left(1+\mathcal{F}_{4,\phi}+{\cal F}_{4,\chi\theta\eta}\right),\nonumber
\end{align}
where $\mathcal{F}_{4,\phi}$ collects all the terms coming from $\Delta H$, and ${\cal F}_{4,\chi\theta\eta}$, those from $\Delta H'$.
We have ignored the determinant and numerical prefactor arising from the Gaussian integration over 
 $\bf\chi$,  $\bf\theta$ and $\bf\eta$, as they are $\xi$-independent and common to all contributions. 

Another relevant point comes from what is known as the `linked cluster theorem' in statistical field theory; for an exposition aimed at a broad audience see chapter 5 of \cite{statFTneu}. The essence is that higher powers of the leading order contributions will arise at higher orders (due to disconnected Feynman diagrams consisting of multiple copies of lower-order diagrams). It is possible to resum all of these corrections in a compact form, which amounts to replacing $1+{\cal F}_{4,\phi}+{\cal F}_{4,\chi\theta\eta}$ in (\ref{eq:saddle_det}) by $e^{{\cal F}_{4,\phi}+{\cal F}_{4,\chi\theta\eta}}$. Thus, once ${\cal F}_{4,\phi}$ and ${\cal F}_{4,\chi\theta\eta}$ have been computed, a wise way to incorporate them into the estimation of  (\ref{Pfull}-\ref{Hprime}) is by writing
\begin{equation}
P(\rho)\propto  \int d\xi \,\,\left(\frac{1}{\phi_0}-2i\xi\phi_0\right)^{-1} e^{-i\xi\rho} e^{\frac{2}{\phi_0}\left(1-\frac{1}{2\kk}\right)+4i\xi\phi_0\left(1+\frac1{2c}\right)+\mathcal{F}_{4,\phi}+\mathcal{F}_{4,\chi\theta\eta}}.
\label{eq:FTcorr1/k}
\end{equation}
This formula provides the $1/\kk$-corrected version of the resistance distance distribution. The rest of this appendix simply reports the computation of ${\cal F}_{4,\phi}$ and ${\cal F}_{4,\chi\theta\eta}$.

\subsection*{Computation of $\mathcal{F}_{4,\phi}$}\label{corr}

We now proceed with evaluating ${\cal F}_{4,\phi}$, whose Feynman diagram representation is 
\beq
\begin{split}
\feynmandiagram [horizontal = a to c]{
    a --[half left] b[dot] -- [half left] c -- [half left] b --[half left] a
    };
\label{butterfly}
\end{split}
\eeq 
We emphasize again that, because of the structure of (\ref{eq:FTcorr1/k}),
we can safely ignore any $\xi$-independent terms in ${\cal F}_{4,\phi}$, as they can be merged into the normalization of (\ref{eq:prho_new}). The quartic terms coming from (\ref{eq:Hexpand}) can be further simplified for the  purposes of computing (\ref{eq:saddle_det}). It is important to keep in mind that the Gaussian measure in (\ref{eq:saddle_det}) is invariant under interchange of $\varphi_1$ and $\varphi_2$, and under arbitrary permutations of $\phi_3,\ldots,\phi_N$. Hence, integrating $(\varphi_1-\phi_j)^4$ or  $(\varphi_2-\phi_j)^4$ with any $j\ge 3$ gives the same result as integrating $(\varphi_1-\varphi_3)$, while integrating $(\phi_i-\phi_j)^4$ with $3\le i<j$ gives the same result as integrating $(\phi_3-\phi_4)^4$. Thus, all the sums in (\ref{eq:Hexpand}) can be eliminated at the cost of appending explicit $N$-dependent factors that count the number of terms, since all the terms in each sum equal each other upon the integration in (\ref{eq:saddle_det}). Once this simplification has been implemented, one can further discard any contributions that vanish at $N\to\infty$. This leads to the following equivalent replacement for  $(\Delta H)_4$, the quartic part of $\Delta H$:
\beq
(\Delta H)_4\to
\frac{\xi^2}{3\phi_0}(4\xi^2\phi_0^4+12i\xi\phi_0^2-3)(\varphi_1-\phi_3)^4
-\frac{(N-5)}{4}\,\kk\,\xi^2(\phi_3-\phi_4)^4.
\label{eq:phi4corr}
\eeq

In general, expanding $(\phi_i-\phi_j)^4=\phi_i^4 - 4\phi_i^3\phi_j + 6\phi_i^2\phi_j^2 - 4\phi_i\phi_j^3 + \phi_j^4$ and applying the standard Gaussian integration leads to
\begin{equation}
\begin{split}
\int d\pmb{\phi}&\left(\sum_i \phi_i\right)e^{-\frac{1}{2}(-M_{ij})\phi_i\phi_j} (\phi_i^4 - 4\phi_i^3\phi_j + 6\phi_i^2\phi_j^2 - 4\phi_i\phi_j^3 + \phi_j^4)\\
&= \frac{(-2\pi)^{\frac{N-1}{2}}}{\sqrt{\pdet{\mathbf{M}}}}\left[3(M^{-1}_{ii})^2 - 4 M^{-1}_{ii}M^{-1}_{ij} + 6M^{-1}_{ii}M^{-1}_{jj}-4M^{-1}_{ij}M^{-1}_{jj}+3(M^{-1}_{jj})^2\right],
\label{eq:quart_expd}
\end{split}
\end{equation}
where ${\bf M}^{-1}$ evidently means the inverse of $\bf M$ in the subspace orthogonal to the null vector $(1,1,1,\ldots)^T$ such that ${\bf M}^{-1}{\bf M}={\bf I}-{\bf 1}/N$.
Evaluating ${\cal F}_{4,\phi}$ amounts to applying this integration formula to \eqref{eq:phi4corr}, and subsequently omitting all terms that are either $\xi$-independent or vanish at $N\to\infty$. While integrating the last term in \eqref{eq:phi4corr}, it is useful to remember that $M_{33}^{-1}=M_{44}^{-1}$ while $M_{34}^{-1}$ is of order $1/N$ and furthermore its leading $1/N$ part is $\xi$-independent. As a result, one obtains
\begin{equation}
{\cal F}_{4,\phi}=\frac{\xi^2}{\phi_0}(4\xi^2\phi_0^4+12i\xi\phi_0^2-3)(M^{-1}_{11} + M^{-1}_{33})^2 +3(N-5)\kk\xi^2 (M^{-1}_{33})^2.
\label{F4M}
\end{equation}

Given the complete description of the eigenvectors and eigenvalues of $\bf M$ under (\ref{eq:Hconstraint}), one can straightforwardly construct ${\bf M}^{-1}$ as
\beq
M^{-1}_{ij} = \frac{1}{a-g}\, e^{1}_ie^{1}_j - \frac{1}{Nf}\, e^{2}_ie^{2}_j + \frac{1}{b-d}\left(\delta_{ij} -e^{0}_ie^{0}_j - e^{1}_ie^{1}_j - e^{2}_ie^{2}_j\right)
\eeq
with ${\bf e}^0=(1,1,1,\ldots)^T/\sqrt{N}$, ${\bf e}^1=(1,-1,0,0,\ldots)^T/\sqrt{2}$, ${\bf e}^2=\sqrt{\frac{N-2}{2N}}(1,1,-\frac{2}{N-2},-\frac{2}{N-2},\ldots)^T$. From this expression, one directly recovers
\beq
M^{-1}_{11}=\frac1{a-g}+O(1/N),\qquad M^{-1}_{33}=\frac{1-1/N}{b-d}+O(1/N^2).
\label{eq:Minverse}
\eeq
Furthermore, $a=a_0+O(1/N)$ and $b=b_0+b_1/N+O(1/N^2)$ with
\beq
a_0 = -4\xi^2\phi_0  -\frac{2i\xi}{\phi_0},\qquad
b_0 = -2i\xi\kk,\qquad
b_1 = -8\xi^2\phi_0-\frac{4i\xi}{\phi_0},
\label{eq:a0b0}
\eeq
while $d$ and $g$ are $O(1/N)$.
We can keep only the contributions of order $1$ in the first term of (\ref{F4M}), meaning that $M^{-1}_{11}\approx1/a_0$ and $M^{-1}_{33}\approx1/b_0$ for the purposes of evaluating this term, while the last term contains an explicit factor of $N$ and we must be careful to retain the contributions of order $1/N$. Note that the contribution of order $N$
in the last term of (\ref{F4M}) is $\xi$-independent and can thus be ignored in the context of (\ref{eq:FTcorr1/k}), while the remaining subleading contribution is of order 1 and $\xi$-dependent. Explicitly,
\beq
(M^{-1}_{33})^2=\frac1{b_0^2}-\frac2{Nb_0^2}-\frac{2}{N}\frac{b_1-dN}{b_0^3}+O(1/N^2).
\eeq
Putting everything together and discarding all terms that vanish at $N\to\infty$ or are $\xi$-independent results in the following evaluation:
\begin{equation}
{\cal F}_{4,\phi}\to\frac{\xi^2}{\phi_0} (4\xi^2\phi_0^4+12i\xi\phi_0^2-3)\left(\frac{1}{a_0}+\frac{1}{b_0}\right)^2 + 3\kk\xi^2 \left( -\frac{2b_1}{b_0^3}\right),
\end{equation}
or explicitly:
\begin{equation}
{\cal F}_{4,\phi}=-\frac{1}{\phi_0\kk^2}\left\{3-6 i\xi  \phi_0^2 -\frac14\left(3-12i\xi\phi_0^2-4\xi^2\phi_0^4\right) \left(1+\frac{\kk\phi_0}{
   1-2i\xi\phi_0^2 }\right)^2\right\}.
\label{eq:F4phi}
\end{equation}

\subsection*{Computation of ${\cal F}_{4,\chi\theta\eta}$}
 
As explained above, only terms in $\Delta H'$ quadratic in $\tilde{\pmb\phi}$ and either quadratic in $\bf\chi$ or bilinear in $\bf\theta$ and $\bf\eta$ contribute nontrivially to ${\cal F}_{4,\chi\theta\eta}$ at order $1/\kk$. In terms of Feynman diagrams, these terms are visualized as
\begin{equation}
\begin{split}
\feynmandiagram [horizontal = a to c]{
    a --[half left] b[dot] -- [scalar,half left] c -- [half left,scalar] b --[half left] a
    };
\hspace{3cm}
\feynmandiagram [horizontal = a to c]{
    a --[half left] b[dot] -- [fermion,half left] c -- [half left,fermion] b --[half left] a
    };
\end{split}
\end{equation}
where the dashed line denotes the $\bf\chi$-propagator, and the line with arrows, the $\bf\theta\eta$-propagator.

In a manner similar to the treatment of $\mathcal{F}_{4,\phi}$, permutation symmetries of the Gaussian measure and the $N\to\infty$ limit let one simplify $\Delta H'$ to obtain
\begin{align}
\Delta H'&\to\frac{ 2i\xi}{\phi_0}(1-2i\xi\phi_0^2) (\varphi_1-\phi_3)^2[(\chi_1-\chi_3)^2+(\theta_1-\theta_3)(\eta_1-\eta_3)]\label{eq:DeltaHprime_expd}\\
&+i\xi\kk(N-5)(\phi_3-\phi_4)^2[(\chi_3-\chi_4)^2+(\theta_3-\theta_4)(\eta_3-\eta_4)].\nonumber
\end{align}
Furthermore, the integration measure of $\pmb\chi$, $\pmb\theta$ and $\pmb\eta$ is fully symmetric under all permutations and decoupled from $\pmb\phi$. Hence, another equivalent replacement is
\beq
\Delta H'\to\left(\frac{ 2i\xi}{\phi_0}(1-2i\xi\phi_0^2) (\varphi_1-\phi_3)^2+i\xi\kk(N-5)(\phi_3-\phi_4)^2\right)[(\chi_1\hspace{-0.7mm}-\hspace{-0.7mm}\chi_2)^2+(\theta_1\hspace{-0.7mm}-\hspace{-0.7mm}\theta_2)(\eta_1\hspace{-0.7mm}-\hspace{-0.7mm}\eta_2)]\label{eq:DeltaHprime_expd}.
\eeq
Since the Gaussian measure in (\ref{eq:FTcorr1/k}) factorizes, the content of the square brackets in the above expression can be treated separately. We first write
\begin{equation}
\frac{\int d\pmb{\chi}\; \delta\left({\textstyle\sum_i \chi_i}\right)e^{-\frac12\sum_{ij} \Pi_{ij}\chi_i\chi_j}(\chi_1-\chi_2)^2}{\int d\pmb{\chi}\; \delta\left({\textstyle\sum_i \chi_i}\right)e^{-\frac12\sum_{ij} \Pi_{ij}\chi_k\chi_l}} = \Pi^{-1}_{11}+\Pi^{-1}_{22}+O(1/N),
\label{eq:coeff_chi}
\end{equation}
where $\mathbf{\Pi} \equiv 2c({\bf I}-{{\bf 1}}/{N})$ is read off (\ref{eq:Hprime}). Similarly,
\begin{equation}
\frac{\int d\pmb\theta d\pmb\eta\,\delta({\textstyle\sum_i \theta_i})\delta({\textstyle\sum_i \eta_i})\, e^{-\frac12\sum_{ij}\Pi_{ij}\theta_i\eta_j}(\theta_1\hspace{-0.7mm}-\hspace{-0.7mm}\theta_2)(\eta_1\hspace{-0.7mm}-\hspace{-0.7mm}\eta_2)}{\int d\pmb\theta d\pmb\eta\,\delta({\textstyle\sum_i \theta_i})\delta({\textstyle\sum_i \eta_i})\,  e^{-\frac12\sum_{ij}\Pi_{ij}\theta_i\eta_j}}=-2\left(\Pi^{-1}_{11}+\Pi^{-1}_{22}+O(1/N)\right)
\label{eq:coeff_te}.
\end{equation}
Then, computing the Gaussian average over $\tilde{\pmb\phi}$ in a manner directly analogous to the evaluation of $\mathcal{F}_{4,\phi}$, we get:
\beq
\mathcal{F}_{4,\xi\theta\eta}=-\left(\Pi^{-1}_{11}+\Pi^{-1}_{22}\right)\left(\frac{ 2i\xi}{\phi_0}(1-2i\xi\phi_0^2)\left(M^{-1}_{11}+M^{-1}_{33}\right) + iN\kk\xi\left(M^{-1}_{33}+M^{-1}_{44}\right)\right).
\eeq
We then recall (\ref{eq:Minverse}-\ref{eq:a0b0}) and approximate $M_{11}^{-1}\approx{1}/{a_0}$ and $M_{33}^{-1}\approx{1}/{b_0}$ while being careful about contributions of order $1/N$ in the term that explicitly involves $N$. We thus arrive at
\begin{equation}
\mathcal{F}_{4,\xi\theta\eta} =-\frac{1}{\kk}\left(\frac{ 2i\xi}{\phi_0}(1-2i\xi\phi_0^2)\left(\frac{1}{a_0}+\frac{1}{b_0}\right) + i\kk\xi\left(\frac{-2b_1}{b_0^2}\right)\right),
\end{equation}
which can be simplified to
\begin{equation}
\mathcal{F}_{4,\xi\theta\eta} = \frac{2 i \phi_0^2 \xi +\kk  \phi_0-1}{\kk ^2 \phi_0 }
\label{eq:F4chi}.
\end{equation}



\begin{thebibliography}{99}

\bibitem{harary}F.~Harary, {\it Graph theory} (CRC Press, 1994).

\bibitem{dist}P.~Miasnikof, A.~Y.~Shestopaloff, L.~Pitsoulis, A.~Ponomarenko and Yu.~Lawryshyn,
{\it Distances on a graph,} in {\it Complex Networks \& Their Applications IX} (Springer, 2021).

\bibitem{geo}J.~Bouttier, P.~Di Francesco and E.~Guitter, {\it Geodesic distance in planar graphs,} 
Nucl.\ Phys.\ B {\bf 663} (2003) 535 \arXiv{cond-mat/0303272} [cond-mat.stat-mech].

\bibitem{resdist1}
P.~G.~Doyle and J.~L.~Snell, {\it Random walks and electric networks} (Mathematical Association of America, 1984) \arXiv{math/0001057} [math.PR].

\bibitem{resdist2}
D.~J.~Klein and M.~Randi\'c, {\it Resistance distance,} J.\ Math.\ Chem. {\bf 12}(1993) 81.

\bibitem{resdist3}
D.~J.~Klein, {\it Resistance-distance sum rules,} Croat.\ Chem.\ Acta {\bf 75} (2002) 633.

\bibitem{resdist4}
W.~Xiao and I.~Gutman, {\it Resistance distance and Laplacian spectrum,} Theor. Chem. Acc. {\bf 110} (2003) 284.

\bibitem{resdist5}
E.~W.~Weisstein, {\it Resistance distance} at MathWorld\\
\href{https://mathworld.wolfram.com/ResistanceDistance.html}{https://mathworld.wolfram.com/ResistanceDistance.html}

\bibitem{commute}
P.~Tetali, {\it Random walks and the effective resistance of networks,}
J.\ Th.\ Prob. {\bf 4} (1991) 101.

\bibitem{lovasz}L.~Lov\'asz, {\it Random walks on graphs: a survey,} in
{\it Combinatorics: Paul Erd\H{o}s is Eighty} v. 2 (1993)
\href{https://web.cs.elte.hu/~lovasz/erdos.pdf}{https://web.cs.elte.hu/$\sim$lovasz/erdos.pdf}.

\bibitem{lostspace1}
U.~von~Luxburg, A.~Radl and M.~Hein, {\it Getting lost in space: large sample analysis of the resistance distance,} Adv.\ Neur.\ Inf.\ Proc.\ Sys.\ {\bf 23} (2010) 2622; {\it Hitting and commute times in large random neighborhood graphs,} J.\ Mach.\ Learn.\ Res. {\bf 15} (2014) 1751.

\bibitem{phys1}F.~Y.~Wu, {\it Theory of resistor networks: the two-point resistance,}
J. Phys. A {\bf 37} (2004) 6653 \arXiv{math-ph/0402038}.

\bibitem{phys2}P.~Van Mieghem, K.~Devriendt and H.~Cetinay,
{\it Pseudoinverse of the Laplacian and best spreader node in a network,}
Phys.\ Rev.\ E {\bf 96} (2017) 032311.

\bibitem{chem1}
N.~Trinajsti\'c, D.~Babi\'c, S.~Nikoli\'c, D.~Plav\v{s}i\'c, D.~Ami\'c and Z.~Mihali\'c, {\it The Laplacian matrix in chemistry,} J. Chem. Inf. Comp. Sci. {\bf 34} (1994) 368.

\bibitem{chem1a}
O. Ivanciuc, {\it QSAR and QSPR molecular descriptors computed from the resistance distance and electrical conductance matrices,} ACH Model. Chem. {\bf 137} (2000) 607.

\bibitem{chem2}D. Babi\'c, D. J. Klein, I. Lukovits, S. Nikoli\'c and N. Trinajsti\'c, {\it Resistance-distance matrix: a computational algorithm and its application,} Int. J. Quant. Chem. {\bf 90} (2001) 166.

\bibitem{chem2a}P.~W.~Fowler, {\it Resistance distances in fullerene graphs,}
Croat. Chem. Acta, 75 (2002) 401.

\bibitem{chem3}K.~Roy, {\it Topological descriptors in drug design and modeling studies}, Molec. Diversity {\bf 8} (2004) 321.

\bibitem{chem4}G. Guillot, R. Leblois, A. Coulon and A.~C. Frantz, {\it Statistical methods in spatial genetics,}
Molec. Ecology {\bf 18} (2009) 4734.

\bibitem{graphinv}Y.~Yang and D.~J.~Klein, {\it Resistance distance-based graph invariants of subdivisions and triangulations of graphs,} Discr. App. Math. {\bf 181} (2015) 260.

\bibitem{cacti1}
J.~Du, G.~Su, J.~Tu and I.~Gutman, {\it The degree resistance distance of cacti}, Discr. App. Math. {\bf 188} (2015) 16.

\bibitem{cacti2}
J.-B.~Liu, W.-R. Wang, Y.-M. Zhang and X.-F. Pan, {\it On degree resistance distance of cacti,} Discr. App. Math. {\bf 203} (2016) 217.

\bibitem{cs1}
D. Liben-Nowell and J. Kleinberg, {\it The link prediction problem for social networks,} in CIKM'2003 Proceedings.

\bibitem{cs2}
D. Zhou and B. Sch\"olkopf, {\it Learning from labeled and unlabeled data using random walks,} in DAGM'2004 Proceedings.

\bibitem{cs3}
M. Saerens, F. Fouss, L. Yen and P. Dupont {\it The principal components analysis of a graph, and its relationships to spectral clustering,} in ECML'2004 Proceedings.

\bibitem{cs4}
L.~Yen, D.~Vanvyve, F.~Wouters, F.~Fouss, M.~Verleysen and M.~Saerens, {\it Clustering using a random walk based distance measure,} in ESANN'2005 Proceedings.

\bibitem{cs5}
M. Herbster and M. Pontil, {\it Prediction on a graph with a perceptron,} in NIPS'2006 Proceedings.

\bibitem{cs6}
C. Avin and G. Ercal, {\it On the cover time and mixing time of random geometric graphs,} Theor. Comput. Sci, {\bf 380} (2007) 2.

\bibitem{cs7}
D. Spielman and N. Srivastava, {\it Graph sparsification by effective resistances,}  in STOC'2008 Proceedings.

\bibitem{cs8}
L.  Yen,  M.  Saerens,  A.  Mantrach  and  M.  Shimbo, {\it A  family  of  dissimilarity  measures between nodes generalizing both the shortest-path and the commute-time distances,}  in SIGKDD'2008 Proceedings.

\bibitem{cs9}
N.~Cesa-Bianchi, C.~Gentile and F.~Vitale, {\it Fast and optimal prediction on a labeled tree,}
in COLT'2009 Proceedings.

\bibitem{cs10}
D.~M.~Wittmann, D.~Schmidl, F. Bl\"ochl and F. J. Theis, {\it Reconstruction of graphs based on random walks,} Theor.\ Comp.\ Sci.\ {\bf 410} (2009) 3826.

\bibitem{cs11}
C.~Cooper and A.~Frieze, {\it The cover time of random geometric graphs,} Rand.\ Struct.\ Alg.\ {\bf 38} (2011) 324.

\bibitem{cs12}
P.~Wills and F.~G.~Meyer, {\it Metrics for graph comparison: a practitioner’s guide,} PLoS ONE 15 (2020) e0228728
\arXiv{1904.07414} [stat.AP].

\bibitem{cs13}A.~Ponomarenko, L.~Pitsoulis and M.~Shamshetdinov,
{\it Overlapping community detection in networks based on link partitioning and partitioning around medoids,} \arXiv{1907.08731} [cs.SI].

\bibitem{vasc1}E.~Katifori, G.~J.~Sz\"oll\H{o}si and M.~O.~Magnasco,
{\it Damage and fluctuations induce loops in optimal transport networks,}
Phys.\ Rev.\ Lett. {\bf 104} (2010) 048704
\arXiv{0906.0006} [physics.bio-ph]

\bibitem{vasc2}T.~Gavrilchenko and E.~Katifori,
{\it Distribution networks achieve uniform perfusion through geometric self-organization,}
\arXiv{2009.04375} [physics.bio-ph].

\bibitem{vasc3}S.~Fancher and E.~Katifori,
{\it Tradeoffs between energy efficiency and mechanical response in fluid flow networks,}
\arXiv{2102.13197} [physics.bio-ph].

\bibitem{vasc4}Y.~Luo, Ch.-L.~Ho, B.~R.~Helliker and E.~Katifori,
{\it Leaf water storage and robustness to intermittent drought: 
a spatially explicit capacitive model for leaf hydraulics,}
\arXiv{2106.08939} [physics.bio-ph].

\bibitem{diam1} T.~\L uczak, {\it Random trees and random graphs,} Rand. Struct. Alg. {\bf 13} (1998) 485.

\bibitem{diam2}A.~K.~Hartmann and M.~M\'ezard, {\it Distribution of diameters for Erd\H{o}s-R\'enyi random graphs,} Phys. Rev. E {\bf 97} (2018) 032128 \arXiv{1710.05680} [cond-mat.dis-nn].

\bibitem{short1}E.~Katzav, M.~Nitzan, D.~ben-Avraham, P.~L.~Krapivsky, R.~K\"uhn, N.~Ross and O.~Biham, {\it Analytical results for the distribution of shortest path lengths in random networks,} EPL {\bf 111} (2015) 26006 \arXiv{1504.00754} [cond-mat.dis-nn].

\bibitem{short2}M.~Nitzan, E.~Katzav, R.~K\"uhn and O.~Biham, {\it Distance distribution in configuration model networks,} Phys. Rev. E {\bf 93} (2016) 062309 \arXiv{1603.04473} [cond-mat.dis-nn].

\bibitem{short3}E.~Katzav, O.~Biham and A.~K.~Hartmann, {\it The distribution of shortest path lengths in subcritical Erd\H{o}s-R\'enyi networks,} Phys. Rev. E {\bf 98} (2018) 012301 \arXiv{1806.05743} [cond-mat.dis-nn].

\bibitem{short4}
A.~D.~Jackson and S.~P.~Patil, {\it Phases of small worlds: a mean field formulation,}\\
\arXiv{2103.14001} [cond-mat.stat-mech].

\bibitem{resdistconc1}
N.~Boumal and X.~Cheng, {\it Concentration of the Kirchhoff index for Erd\H{o}s-R\'enyi graphs,}
Sys.\ Contr.\ Lett. {\bf 74} (2014) 74 \arXiv{1307.6398} [cs.IT].

\bibitem{resdistconc2}J.~Sylvester,
{\it Random walk hitting times and effective resistance in sparsely connected Erd\H{o}s-R\'enyi random graphs,} J. Graph Th. {\bf 96} (2021) 44 \arXiv{1612.00731} [math.CO].

\bibitem{lostspace2}
M.~Alamgir and U.~von Luxburg, {\it Phase transition in the family of p-resistances,} in {\it Proceedings of the 24th International Conference on Neural Information Processing Systems} (2011) p.~379.

\bibitem{randres}I.~Benjamini and R.~Rossignol,
{\it Submean variance bound for effective resistance of random electric networks,}
Comm. Math. Phys. {\bf 280} (2008) 445
\arXiv{math/0610393} [math.PR].

\bibitem{minimize}A.~Gosh, S.~Boyd and A.~Saberi, {\it Minimizing effective resistance
of a graph,} SIAM Rev. {\bf 50} (2008) 37.

\bibitem{statFT}E.~Br\'ezin, {\it Introduction to statistical field theory} (Cambridge,  2010).

\bibitem{statFTneu}M.~Helias and D.~Dahmen, {\it Statistical field theory for neural networks} (Springer, 2020).

\bibitem{largeN}M.~Moshe and J.~Zinn-Justin,
{\it Quantum field theory in the large N limit: a review,}
Phys. Rept. \textbf{385} (2003) 69
\arXiv{hep-th/0306133}.

\bibitem{PN}J.~Park and M. E. J. Newman, {\it Solution of the two-star model of a network}, Phys.\ Rev.\ E {\bf 70} (2004) 066146 \arXiv{cond-mat/0405457}.

\bibitem{spectrum}F.~L.~Metz, G.~Parisi and L.~Leuzzi, {\it Finite size correction to the spectrum of regular random graphs: an analytical solution,} Phys. Rev. E {\bf 90} (2014) 052109 \arXiv{1403.2582} [cond-mat.dis-nn].

\bibitem{AC}A.~Annibale and O.~T.~Courtney, {\it The two-star model: exact solution in the sparse regime and condensation transition}, J.\ Phys.\ A {\bf 48} (2015) 365001 \arXiv{1504.06458} [cond-mat.dis-nn].

\bibitem{corr}M.~Bolfe, F.~L.~Metz, E.~Guzm\'an-Gonz\'alez and I.~P\'erez Castillo, {\it Analytic solution of the two-star model with correlated degrees,} Phys.\ Rev.\ E {\bf 104} (2021)  014147 \arXiv{2102.09629} [cond-mat.stat-mech].

\bibitem{BG}M.~Bauer and O.~Golinelli, {\it Random incidence matrices: moments of the spectral density,}
J. Stat. Phys. {\bf 103} (2001) 301 \arXiv{cond-mat/0007127}.

\bibitem{newman}M.~Newman, {\it Networks: an introduction} (Oxford, 2010).

\bibitem{MF}A.~D.~Mirlin and Y.~V.~Fyodorov, {\it Universality of level correlation function of sparse random matrices,} J.\ Phys.\ A {\bf 24} (1991) 2273.

\bibitem{euclRM}
M.~M\'ezard, G.~Parisi and A.~Zee,
{\it Spectra of Euclidean random matrices,}
Nucl. Phys. B \textbf{559} (1999) 689
\arXiv{cond-mat/9906135}.

\bibitem{ecological}
 Y.~Krumbeck, Q.~Yang, G.~W.~A.~Constable and T.~Rogers,
{\it Fluctuation spectra of large random dynamical systems reveal hidden structure in ecological networks,}
Nature\ Comm. {\bf 12} (2021) 3625 \arXiv{2011.05140} [q-bio.PE].

\bibitem{span1}
R.~Lyons, {\it Asymptotic enumeration of spanning trees,} Comb. Prob. Comp. {\bf 14} (2005) 491
\arXiv{math/0212165} [math.CO].

\bibitem{span2}
R.~Lyons, R.~Peled and O.~Schramm, {\it Growth of the number of spanning trees of the Erd\H{o}s-R\'enyi giant component,} Comb. Prob. Comp. {\bf 17} (2008) 711 \arXiv{0711.1893} [math.PR].

\bibitem{efetov}K.~Efetov, {\it Supersymmetry in disorder and chaos} (Cambridge, 1996).

\bibitem{fyodorov}Y.~V.~Fyodorov, {\it Complexity of random energy landscapes, glass transition and absolute value of spectral determinant of random matrices,}
Phys. Rev. Lett. {\bf 92} (2004) 240601 \arXiv{cond-mat/0401287} [cond-mat.dis-nn]. 

\bibitem{FN}
Y.~V.~Fyodorov and A.~Nock, {\it On random matrix averages involving half-integer powers of GOE characteristic polynomials,} J. Stat. Phys. {\bf 159} (2015) 731 \arXiv{1410.5645} [math-ph].

\bibitem{lambertW}
R.~M.~Corless, G.~H.~Gonnet, D.~E.~G.~Hare, D.~J.~Jeffrey and D.~E.~Knuth,
{\it On the Lambert W function,} Adv. Comp. Math. {\bf 5} (1996) 329.

\bibitem{wong}R.~Wong, {\it Asymptotic approximations of integrals} (SIAM, 2001), section VII.4.

\bibitem{overlap}
H. F. Inman and E. L. Bradley, {\it The overlapping coefficient as a measure of agreement between probability distributions and point estimation of the overlap of two normal densities,} Comm. Stat. Th. Meth. {\bf 18} (1989) 385.

\bibitem{hetero}
R.~Kuehn and T.~Rogers,
{\it Heterogeneous micro-structure of percolation in sparse networks,}
Eur.\ Phys.\ Lett. {\bf 118} (2017) 68003  \arXiv{1703.06740} [cond-mat.stat-mech].

\end{thebibliography}
\end{document}